\documentclass[twocolumn,aps,prd,amsmath,floats,floatfix,nofootinbib]{revtex4-2}

\usepackage[utf8]{inputenc}  % PDFLaTeX 支持 UTF-8
\usepackage[T1]{fontenc}  %for non-ascii encoding
\usepackage[normalem]{ulem} % strike out without modifying emph
\usepackage{amsmath,amssymb}
\usepackage{graphicx}
\usepackage{tikz}       % use for overlay
\usepackage{dcolumn}
\usepackage{diagbox}
\usepackage{bm}
\usepackage{bbm}
\usepackage{xcolor}
\usepackage[multiple]{footmisc}
\usepackage[colorlinks,linkcolor=magenta,anchorcolor=cyan,citecolor=blue]{hyperref}

\def\be{\begin{equation}}
\def\ee{\end{equation}}
\def\ba{\begin{eqnarray}}
\def\ea{\end{eqnarray}}

     \def\i {\iota}                    
              \def\.{\cdot}

\begin{document}

\title{Black hole destabilization via trapped quasi-normal modes}

\author{Hsu-Wen Chiang}
%\email{jiangxw[at]sustech.edu.cn}
%\email{b98202036[at]ntu.edu.tw}

\author{Sebastian Garcia-Saenz}
%\email{sgarciasaenz[at]sustech.edu.cn}

\author{Aofei Sang}
%\email{12331027[at]mail.sustech.edu.cn}

\affiliation{Department of Physics, Southern University of Science and Technology, Shenzhen 518055, China}

%\date{\today}

\begin{abstract}
In the presence of non-minimal gravitational couplings, matter field perturbations on a static black hole spacetime may develop unphysical poles in their linearized equations. Physical solutions confined in the domain between the event horizon and a pole satisfy a boundary value problem, although with boundary conditions which are different from standard quasi-normal modes. We refer to them as ``trapped quasi-normal modes''. Focusing on a Schwarzschild black hole in Einstein-Proca theory, we find that trapped quasi-normal modes accurately capture the behavior of perturbations under time evolution. In particular, axial-vector modes are unstable, with a growth rate that increases with multipole number. More interestingly, we uncover a new instability that affects monopole perturbations. These results confirm the existence of a novel destabilization mechanism of black holes by non-minimally coupled vector fields, with potential implications to well-studied models of modified gravity and cosmology based on vector particles.

\end{abstract}

\maketitle

\section{Introduction}

The response of black holes to perturbations provides key information for their characterization, and is as such an essential tool in the effort of testing gravity and new physics, for instance the existence of yet-undiscovered light particles, by experimental probes of black holes \cite{Binnington:2009bb,Brito:2014wla,Krishnendu:2017shb,Baumann:2018vus,Cardoso:2019rvt,Brito:2025ojt}. Among the different types of responses, quasi-normal modes (QNMs) are of prime interest, being directly measurable, in the case of gravitational perturbations, in the ring-down phase of a merger event \cite{Dreyer:2003bv,Berti:2005ys,Ferrari:2007dd,Berti:2016lat,LIGOScientific:2021sio}. Furthermore, QNMs also serve to diagnose novel physical effects, in particular instabilities such as those that characterize the superradiance and scalarization mechanisms (see \cite{Brito:2015oca,Doneva:2022ewd} for reviews). 

Mathematically, QNMs are defined as solutions of a specific boundary value problem in the domain exterior to the black hole event horizon (see \cite{Kokkotas:1999bd,Berti:2009kk,Konoplya:2011qq,Jansen:2017oag} for reviews). On the other hand, it has been recently observed that matter fields which couple non-minimally with gravity may develop poles in their equations of motion, at least in the linearized approximation \cite{Garcia-Saenz:2022wsl} (see also \cite{Chen:2010qf,Jusufi:2020odz}). These poles need not be considered as physical as they arise from a perturbative treatment, but nevertheless they do affect the defining boundary value problem of QNMs. Indeed, as we will see, regularity of the perturbation at a pole's location results in an additional boundary condition which prevents one from constructing global solutions with a single QNM spectrum.

On the physical side, the issue of non-minimal gravitational couplings has been argued to be linked to a novel destabilization mechanism of black holes \cite{Garcia-Saenz:2021uyv} (see also \cite{BeltranJimenez:2013btb} for earlier work). This conclusion was drawn from the study of the Einstein-Proca theory describing a massive vector field coupled to gravity, a set-up motivated by a wealth of recent results related to the question of stability of the Proca system \cite{Ramazanoglu:2017xbl,Silva:2021jya,Coates:2022nif,Coates:2022qia,Clough:2022ygm,Coates:2023dmz,Capanelli:2024pzd,Annulli:2019fzq,Kase:2020yhw,Minamitsuji:2020pak,Pizzuti:2023eyt,Ye:2024pyy,Rubio:2024ryv,DeFelice:2025ykh}. Yet, in the non-minimally coupled case, previous analyses were restricted to a short-wavelength approximation, which is oblivious to boundary conditions and is therefore not expected to fully capture the dynamics of perturbations under time evolution, and in particular may not be an accurate tool to diagnose instabilities.

Our aim in this paper is to perform an exact treatment of QNMs in the Einstein-Proca theory with a particular choice of non-minimal coupling, and focusing for simplicity on a Schwarzschild black hole background. Based on both numerical and analytical results, we confirm the existence of instabilities affecting perturbations in the region between the event horizon and the pole, which we refer to as ``trapped QNMs''. We also show that, although distinct from standard QNMs, trapped QNMs also accurately characterize the salient features of the evolution of perturbations in the time domain. Our results for the QNM spectrum show a non-trivial dependence on the model parameters. In particular, we discover a new instability affecting specifically monopole perturbations. 

%-------------------------------
%-------------------------------

\section{Non-minimally coupled Einstein-Proca theory}

Our set-up is the action of Einstein-Proca theory including non-minimal gravitational couplings for the vector field:
\begin{equation} \label{eq:Lagrangian}
\begin{aligned}
S[g,A]&=\int d^4x\sqrt{-g}\bigg[\frac{M_{\rm Pl}^2}{2}R-\frac{1}{4}F^{\mu\nu}F_{\mu\nu}-\frac{\mu^2}{2}A^{\mu}A_{\mu} \\
&\quad +\frac{\alpha}{4}\widetilde{R}^{\mu\nu\rho\sigma}F_{\mu\nu}F_{\rho\sigma}+\beta G^{\mu\nu}A_{\mu}A_{\nu}\bigg] \,.
\end{aligned}
\end{equation}
Here $F_{\mu\nu}$ is the Abelian field strength, $G_{\mu\nu}$ is the Einstein tensor, and $\widetilde{R}^{\mu\nu\rho\sigma}\equiv \frac{1}{4}\epsilon^{\mu\nu\mu'\nu'}\epsilon^{\rho\sigma\rho'\sigma'}R_{\mu'\nu'\rho'\sigma'}$ is the double-dual Riemann tensor. The scales $M_{\rm Pl}$ and $\mu$ are, respectively, the Planck and Proca field masses, while the non-minimal coupling constants are denoted by $\alpha$ and $\beta$, respectively of mass-dimension $-2$ and $0$.

The above choice of non-minimal couplings is of course not the only one. However it may be motivated from the starting point of the Generalized Proca theory \cite{Tasinato:2014eka,Heisenberg:2014rta}, since it may be shown that the most general consistent linearization (about the trivial state $A_{\mu}=0$) of this theory is given by the result \eqref{eq:Lagrangian}. These couplings may be also shown to be unique under some additional assumptions  \cite{Horndeski:1976gi,Garcia-Saenz:2022wsl}. 

In this work we focus on a Schwarzschild black hole and work at linear order in vector field perturbations. As we consider a Ricci-flat spacetime and neglect backreaction, the second non-minimal coupling (proportional to $\beta$) in \eqref{eq:Lagrangian} plays no role, while the first one simplifies to $\widetilde{R}^{\mu\nu\rho\sigma}F_{\mu\nu}F_{\rho\sigma}=-R^{\mu\nu\rho\sigma}F_{\mu\nu}F_{\rho\sigma}$ ($R_{\mu\nu}=0$). The equation of motion for the vector is then
\begin{equation} \label{eq:covariant eom}
\nabla_\mu F^{\mu\nu} - \mu^2 A^\nu + \alpha R^{\mu\nu\rho\sigma} \nabla_\mu F_{\rho\sigma}=0 \,,
\end{equation}
from which the Lorenz constraint follows~\footnote{This is easier to derive before one simplifies the equation of motion, i.e.~$\nabla_\mu F^{\mu\nu}- \mu^2A^\nu-\alpha\nabla_{\mu}\left( \tilde{R}^{\mu\nu\rho\sigma}F_{\rho\sigma}\right)=0$. It is then immediate to see that the divergence of the last term vanishes identically.}: $\nabla_{\mu}A^{\mu}=0$. In our analysis we will also consider the massless limit, $\mu=0$, as a particular case. In this situation the Lorenz constraint is not a consequence of the field equations but may be imposed as a gauge condition.

We use the standard parametrization for the Schwarzschild line element and employ units with the Schwarzschild radius set to one. The Proca field is decomposed into vector spherical harmonics as $A_\mu = \frac{1}{r} \sum_{i=1}^{4} \sum_{l,m} u_i^{lm}(t,r) Z_\mu^{(i)lm}(\theta,\phi)$, and we refer the reader to \cite{Garcia-Saenz:2022wsl} for our conventions. Notice that we frequently omit the labels $l,m$ on the mode functions $u_i^{lm}$.

Monopole perturbations defined by $l=0$ are distinguished in that the functions $u_3$ and $u_4$ do not enter the spherical harmonic expansion. The constraint Lorenz then relates the remaining variables, so eventually one obtains a single dynamical equation,
\be \label{eq:modeequuM}
\mathcal{D} u_M-\frac{f}{P_-}\left[\mu^2+P_-\left(\frac{2}{r^2}-\frac{3}{r^3}\right)\right]u_M=0 \,,
\ee
for the variable $u_M\equiv u_2^{00}$. Here we have introduced the notation $\mathcal{D}\equiv -\frac{\partial^2}{\partial t^2}+\frac{\partial^2}{\partial r_{\ast}^2}$, where $r_\ast$ is the tortoise coordinate (i.e.\ $dr_\ast=dr/f(r)$, $f(r)\equiv 1-1/r$), and
\be \label{eq:Pminus}
P_-(r) \equiv 1-\frac{r_-^3}{r^3} \,,\qquad r_-\equiv (-2\alpha)^{1/3} \,.
\ee
As anticipated, the mode equation \eqref{eq:modeequuM} features a real pole at $r=r_-$, in addition to the usual poles at $r=0,1,\infty$. If $r_-<1$, this pole is ``hidden'' inside the event horizon and therefore does not qualitatively affect the boundary value problem in the physical domain $r\in (1,\infty)$. On the other hand, we expect substantially novel effects when $r_->1$ (equivalently $\alpha<-1/2$).

The existence of an additional pole in the mode equations persists for higher multipoles \cite{Garcia-Saenz:2022wsl}. For simplicity, in this paper we restrict our attention to axial-vector perturbations characterized by the function $u_4$. Notice that this variable is not constrained by the Lorenz condition and, in the massless case, is automatically gauge-invariant. The mode equation may be expressed as
\be\begin{aligned} \label{eq:axialequ}
\,&\mathcal{D} u_A-\frac{f}{P_+}\bigg[\mu^2 +\frac{l(l+1)P_-}{r^2} \\
& +\frac{9}{4r^2}\left(1-\frac{1}{P_+}\right)\left(f+\left(\frac{5}{3}-\frac{7}{3r}\right)P_+\right)\bigg]u_A=0 \,,
\end{aligned}\ee
with the notation $u_A\equiv P_+^{1/2}u_4$, and where
\be \label{Pplus}
P_+(r)\equiv 1-\frac{r_+^3}{r^3} \,,\qquad r_+\equiv \alpha^{1/3} \,.
\ee
Thus the axial mode equation exhibits an additional real pole at $r=r_+$, which lies in the physical domain when $r_+>1$ (equivalently $\alpha>1$). Notice, incidentally, that $r=r_+$ is a double pole, unlike in the monopole case.

Our goal for the reminder of the paper will be to study the solutions of Eqs.\ \eqref{eq:modeequuM} and \eqref{eq:axialequ} under the assumption that the poles lie ``outside'' the black hole horizon, thus qualitatively affecting the boundary conditions.

%-------------------------------
%-------------------------------

\section{Monopole perturbations}

We Fourier-transform, $u_M(t,r)=e^{-i\omega t} u_M(r)$, to obtain from \eqref{eq:modeequuM} a second-order ODE. As explained, the equation contains three poles of physical relevance, and we find the following asymptotic expressions valid in the vicinity of each:
\be \label{eq:asymptotic_mono}
u_M=\begin{cases}
A_{\rm in}e^{-i\omega r_\ast}+A_{\rm out}e^{i\omega r_\ast}\,, & r\to1 \,,\\
B_1 (r - r_-) \\
+ B_2 \left( 1 + c (r - r_-) \log |r-r_-| \right) \,, & r\to r_- \,,\\
C_{\rm in} e^{-\sqrt{\mu^2-\omega^2} r_\ast}+C_{\rm out}e^{ \sqrt{\mu^2-\omega^2} r_\ast} \,, & r\to \infty \,,
\end{cases}
\ee
where $c\equiv\frac{\mu^2r_-^3}{3\left(r_- - 1 \right)}$~\footnote{The \textit{leading} asymptotic expression for $u_M$ is the same as one approaches $r_-$ from the left or right, but this equality does not hold for the subleading terms in the series. The same comment applies to the axial mode in the next section.}. QNMs are defined by the boundary conditions corresponding to infalling wave at the horizon and outgoing wave at spatial infinity, i.e.\ $A_{\rm out}=0$ and $C_{\rm in}=0$, where we assume $\mu<|\omega|$ in order to have the correct oscillatory behavior at infinity (otherwise the solutions correspond to quasi-bound states). Unlike in the standard situation, here these conditions do not suffice to fix the boundary value problem due to the pole at $r=r_-$. The additional physical input is that observable quantities must be finite at and in the vicinity of the pole~\footnote{We recall that QNMs are guaranteed to be regular at the event horizon, as may be seen explicitly by using hyperboloidal coordinates; see \cite{PanossoMacedo:2024nkw} for a review.}. This implies $B_2=0$. Indeed, a calculation of the electric field $F_{tr}$ results in a divergent expression, $\propto\log|r-r_-|$, unless this choice is made.

Having determined the boundary conditions, the mode equation \eqref{eq:modeequuM} may be solved in each of the regions $r\in(1,r_-)$ (``inside the pole'') and $r\in(r_-,\infty)$ (``outside the pole'') in the frequency domain, where one seeks to determine the spectrum of eigenvalues $\omega$. We also study Eq.\ \eqref{eq:modeequuM} in the time domain, where the PDE is solved for $u_M(t,r)$ given an initial profile $u_M(0,r)$ and suitable boundary conditions. For concreteness we consider a simple Gaussian form as the initial condition: $u(0,r_\ast)=\exp\left[-\left(\frac{r_\ast-a}{\Delta}\right)^2\right]$, $\dot{u}(0,r_\ast)=0$. We have verified that our results are largely insensitive to the parameters $a$ and $\Delta$ that define the initial waveform, provided they are chosen such that the form has zero amplitude, within numerical precision, at each pole (i.e.\ vanishing Dirichlet boundary conditions). Moreover, we have also considered a more general class of initial conditions, again with consistent results. We refer the reader to the Supplemental Material for details, including explanations on the spectral and time domain methods used in our numerical calculations.

We first study the boundary value problem in the ``interior'' of the pole. Spectral analysis reveals that the fundamental mode has a purely imaginary frequency $\omega_{\rm int}$. Such fully damped modes are also found in the standard QNM spectrum of gravitational perturbations \cite{Leaver:1985ax}, although there this phenomenon happens for relatively high overtones and is known to be related to the so-called ``algebraically special'' mode \cite{Chandrasekhar:1984mgh}. Fully damped modes are also characteristic of asymptotically de Sitter black holes \cite{Cardoso:2017soq,Jansen:2017oag,Konoplya:2022xid}, which might suggest an analogy between trapped QNMs and QNMs of spacetimes with two horizons. The present case is however clearly distinct since higher overtones are not purely imaginary. Moreover, de Sitter black holes have a branch of QNMs which approach the purely damped modes of empty de Sitter spacetime in the limit of vanishing black hole mass. In contrast, here the pole $r_-$ also goes to zero if one lets the Schwarzschild radius go to zero, so that there is no equivalent of ``empty de Sitter limit'' for trapped QNMs.

\begin{figure}[h]
\centering
\includegraphics[width=0.47\textwidth]{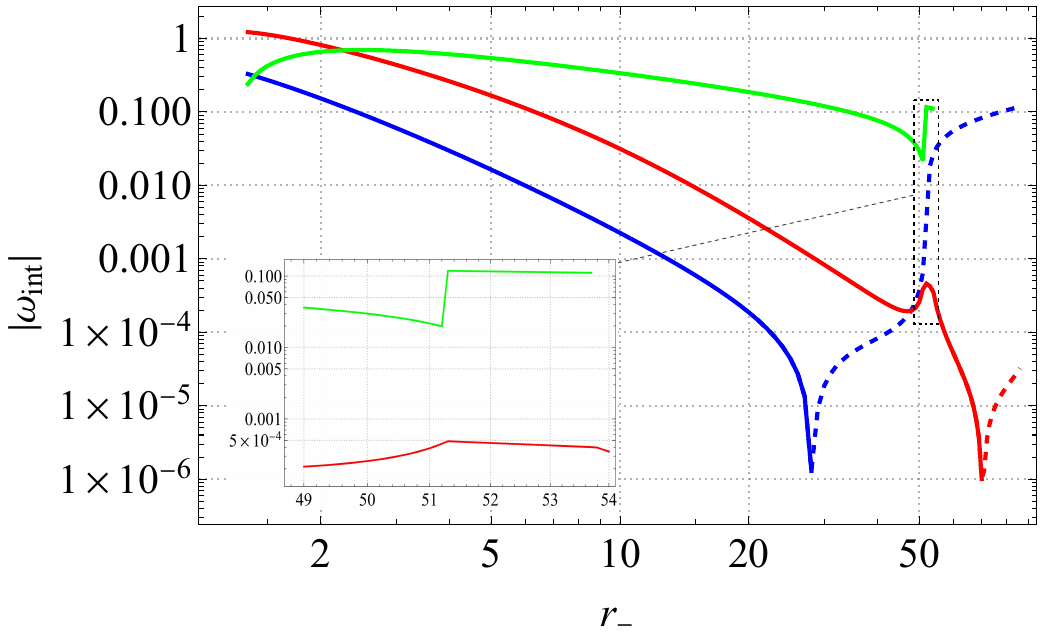}
\caption{Dependence of the ``interior'' frequency $\omega_{\rm int}$ of monopole perturbations on the value of $r_-$, choosing $\mu=1/10$ for the Proca mass. The curves show the imaginary part (blue) of the fundamental mode as well as the real (green) and imaginary (red) parts of the first overtone mode. Solid curved indicate negative (stable) ${\rm Im}\,\omega_{\rm int}$, dashed indicate positive (unstable) ${\rm Im}\,\omega_{\rm int}$.}
\label{fig:mono-im}
\end{figure}

More remarkable is the existence of unstable QNMs, which is seen to occur for relatively large values of $r_-$, as ${\rm Im}\,\omega_{\rm int}$ changes sign and becomes positive. This is illustrated in Fig.\ \ref{fig:mono-im}, where we show the fundamental mode and first overtone for a particular value of the Proca mass $\mu$. In fact, the onset of the instability appears to be controlled by the dimensionless combination $\mu r_-$. This may be understood from inspection of the effective potential in Eq.\ \eqref{eq:modeequuM}, which is approximated by
\begin{equation} \label{eq:VM}
V_M\simeq \frac{(\mu r_-)^2\left(1+1/r_-\right)}{3(r/r_--1)} \,,
\end{equation}
in the vicinity of the pole. Thus $V_M$ is negative in the pole interior, with a ``well'' whose depth is characterized by $\mu r_-$. Strictly speaking, of course, the ``well'' is infinitely deep, however one expects the number of bound states supported but the potential to increase as $\mu r_-$ increases. Fig.\ \ref{fig:mono-im} shows that this is precisely the case, as higher overtones become unstable only for sufficiently large values of $\mu r_-$. What is perhaps unexpected is that there exists a critical value, $\mu r_-\approx 2.8$, below which no unstable modes exist. As suggested by \eqref{eq:VM}, and as verified numerically, this value is approximately universal, independent of $\mu$ and $r_-$, provided $r_-$ is sufficiently large. We also mention that unstable monopole modes are not predicted by the short-wavelength analysis employed in \cite{Garcia-Saenz:2021uyv}, highlighting the fact that this approximation may fail to diagnose instabilities.

Another noteworthy property is the non-trivial dependence of $\omega_{\rm int}$ on the pole distance. We particularly remark on the presence of ``nodes'' in the curves shown in Fig.\ \ref{fig:mono-im} at which their slopes are discontinuous. In some cases this is attributed to a ``crossing'' of modes, e.g.\ when the first and second overtones (the latter not included in Fig.\ \ref{fig:mono-im}) exchange their roles. This explains, in particular, why the real part of the first overtone jumps discontinuously from a finite value to zero: there is another branch of QNMs which at this point takes on the role of first overtone, and this mode is purely damped. Another possibility is that a node corresponds to a bifurcation point, i.e.\ a point where two QNM branches merge into a single one. This is in fact the case for the node at $r_-\simeq 51.2$ in Fig.\ \ref{fig:mono-im}, although again the other branch is not shown as it corresponds to a higher overtone. We mention that the phenomena of QNM crossing and bifurcation have been observed in other systems \cite{Miranda:2008vb,Yang:2012pj,Kokkotas:2015uma,Aragon:2020tvq,Dias:2021yju,Wang:2021upj,Silva:2024ffz,Cavalcante:2024swt,Li:2024npg}; see also \cite{inprep1} for further details.

\begin{figure}[h]
\centering
\includegraphics[width=0.47\textwidth]{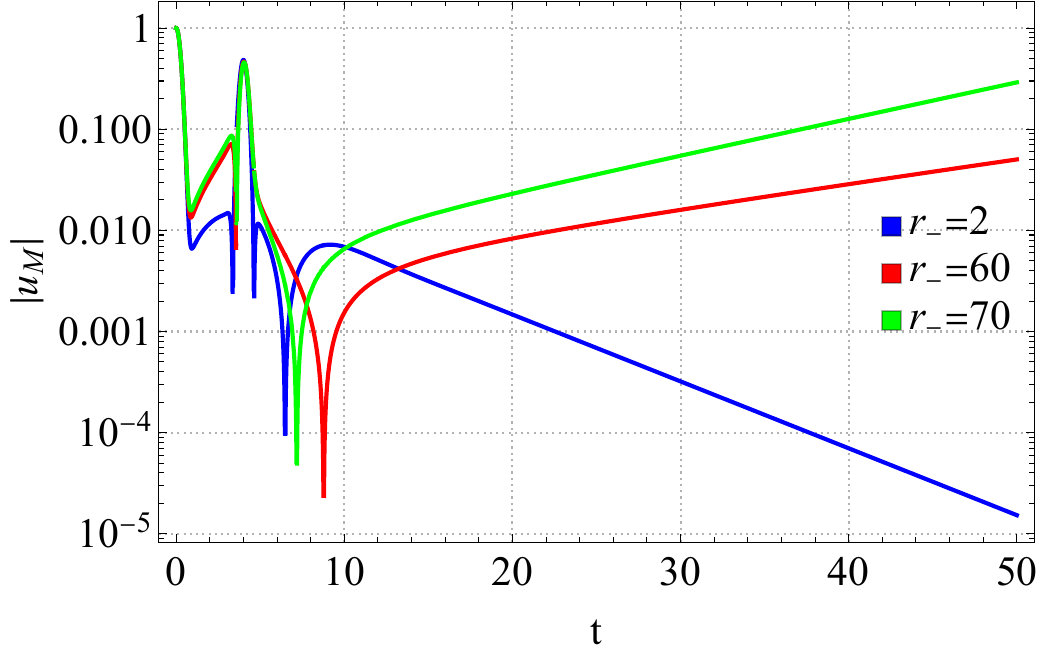}
\caption{Time evolution of $u_M$ inside the pole. The parameter setting is $\mu=1/10$, $a=-2$ and $\Delta=1/3$, with varying values of $r_-$. The mode function is evaluated $r_\ast=a$, where the tortoise coordinate is normalized such that the pole is at $r_\ast=0$.}
\label{fig:mono-1}
\end{figure}

The previous conclusions may be corroborated through a time-domain analysis. Focusing for simplicity on the fundamental mode, we display in Fig.\ \ref{fig:mono-1} the time evolution of the mode function $u_M$ for various values of $r_-$. At late times, after the interactions of the waveform with the pole have subsided, the curve shows a clear exponential behavior, either decaying or growing, and measurement of the slope shows in each case perfect agreement with the spectral method calculation, cf.\ Table \ref{tab:table1}. In particular, we find no evidence of a power-law tail, which is again in analogy with the situation of asymptotically de Sitter black holes \cite{Brady:1996za,Brady:1999wd,Molina:2003dc}.

Considering next the exterior domain $r\in(r_-,\infty)$, our numerical results derived from the spectral and time evolution analyses show no evidence of instabilities (see the Supplemental Material), in accordance with the intuition derived from the form of the effective potential $V_M$, which is strictly positive outside the pole. Moreover, purely damped modes are absent in the spectrum, in contrast to what we observed in the interior domain. We thus conclude, as anticipated, that interior and exterior QNM spectra are different. We have also computed the fundamental mode frequency from the time domain profile using the Prony method. As shown in Table \ref{tab:table1}, the results of both methods agree very well. In addition to the QNM behavior at intermediate timescales, the curve now also exhibits a late-time oscillating tail characterized by the mass scale $\mu$, as is typical of massive fields in asymptotically flat spacetimes \cite{Koyama:2000hj,Koyama:2001ee,Zhidenko:2006rs}.

\begin{table}
\begin{center}
\begin{tabular}{|c|c|c|}
\hline
$\mu=1/10$, $r_-=2$ & $\omega_{\rm int}$ & $\omega_{\rm ext}$ \\ \hline
Spectral method &  $-0.1523 i$  & $\pm0.1668-0.2869 i$ \\ \hline
Prony method &  $-0.1523 i$ &  $\pm0.1664-0.2865 i$ \\ \hline
\end{tabular}
\end{center}
\caption{Comparison of results for the fundamental QNM frequency of monopole perturbations, using a particular parameter setting, for the spectral and Prony methods.}
\label{tab:table1}
\end{table}

The stability of the QNM spectrum in the exterior region is in agreement with the findings of Ref.\ \cite{Garcia-Saenz:2022wsl}, which assumed that the pole was located inside the black hole horizon. Similarly, the analysis of \cite{Garcia-Saenz:2021uyv} concluded, based on dispersion relations derived in a short-wavelength approximation, that all modes are stable outside the pole. On the other hand, the dispersion relation of monopole perturbations was shown to be stable also \textit{inside} the pole. This demonstrates, very notably, that the short-wavelength approximation is in general inconclusive and, in particular, may miss the existence of instabilities.

%-------------------------------
%-------------------------------

\section{Axial-vector perturbations}

We turn to the set-up where $r_+>1$, such that the axial mode equation \eqref{eq:axialequ} exhibits a pole in the physical domain. We find the following asymptotic expressions valid, respectively, in the vicinity of $r=1,r_+,\infty\,$:
\be \label{eq:asymptotic_axial}
u_A=\begin{cases}
A_{\rm in}e^{-i\omega r_\ast}+A_{\rm out}e^{i\omega r_\ast}\,, & r\to1 \,,\\
(r-r_+)^{1/2}(B_1+B_2\log|r-r_+|) \,, & r\to r_+ \,,\\
C_{\rm in} e^{-\sqrt{\mu^2-\omega^2} r_\ast}+C_{\rm out}e^{ \sqrt{\mu^2-\omega^2} r_\ast} \,, & r\to \infty \,.
\end{cases}
\ee
It may be checked that the magnetic field components $F_{r\theta}$ and $F_{r\phi}$ are singular at $r=r_+$ unless $B_2=0$. Thus the boundary value problem is determined by the conditions $A_{\rm out}=B_2=C_{\rm in}=0$ which define physical QNM solutions.

\begin{figure}[h]
\centering
\includegraphics[width=0.47\textwidth]{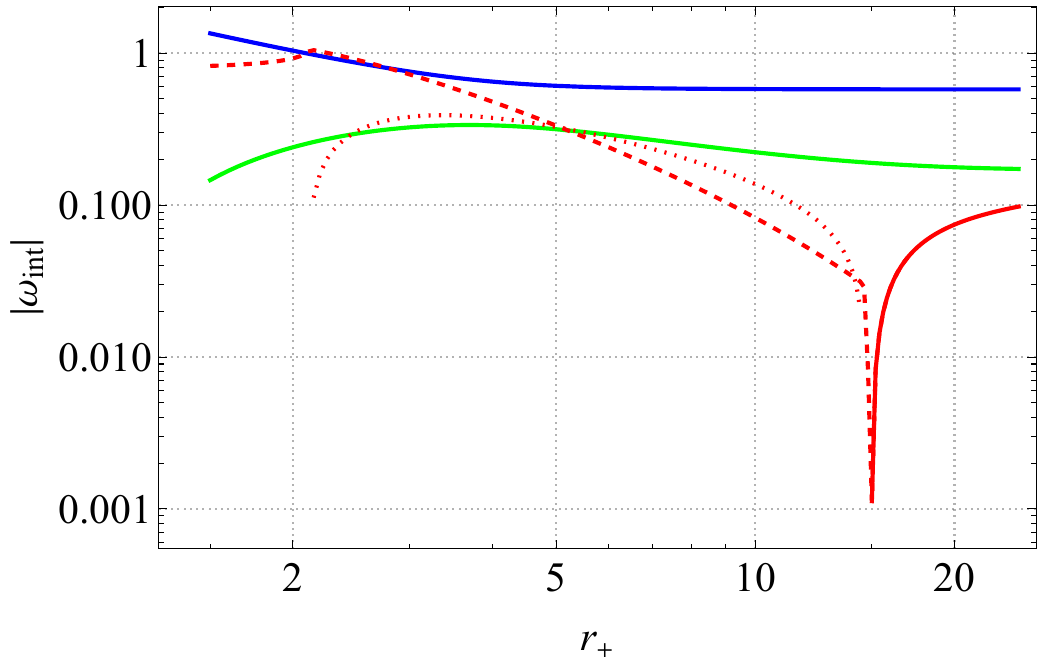}
\caption{Dependence of the ``interior'' frequency $\omega_{\rm int}$ of axial-vector perturbations on the value of $r_+$, with $l=1$ and $\mu=1/10$. The curves show the fundamental (blue), first overtone (green) and second overtone (red) frequencies. Solid curves indicate positive (unstable) ${\rm Im}\,\omega_{\rm int}$, dashed indicates negative (stable) ${\rm Im}\,\omega_{\rm int}$, and dotted indicates ${\rm Re}\,\omega_{\rm int}$ (zero for the fundamental mode and first overtone).}
\label{fig:axial omega}
\end{figure}

Our study of the mode equation in frequency space shows that the fundamental and first overtone modes are unstable in the interior region, for all values of the pole distance, multipole number $l$ and Proca mass $\mu$, at least in the range $\mu\lesssim\mathcal{O}(1)$. On the other hand, higher overtones display the phenomena of mode crossing and bifurcation as functions of $r_+$, similarly to the monopole setting. For instance, using the particular parameter choice of Fig.\ \ref{fig:axial omega}, we find that the second overtone is purely damped for $r_+\lesssim 2.1$, then develops a real part in the range $2.1\lesssim r_+\lesssim 14$, then turns purely damped again before becoming unstable for $r_+\gtrsim 15$. Similarly to the monopole sector, the number of unstable modes increases as $r_+$ becomes large, an observation which is confirmed by an analytical approximation for the QNM frequencies; see the Supplemental Material.

\begin{figure}[t]
\centering
\includegraphics[width=0.48\textwidth]{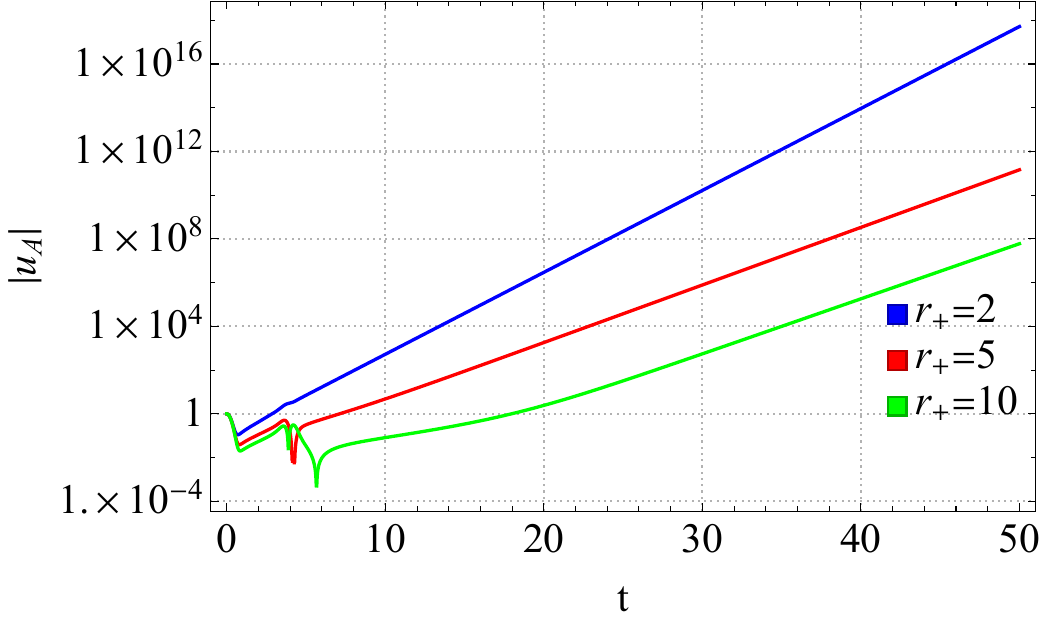}
\caption{Time evolution of $u_A$ inside the pole. The parameter setting is $\mu=1/10$, $l=1$, $a=-2$ and $\Delta=1/3$, with varying values of $r_+$. The mode function is evaluated $r_\ast=a$.}
\label{fig:axial time}
\end{figure}

\begin{figure}[t]
\centering
\includegraphics[width=0.46\textwidth]{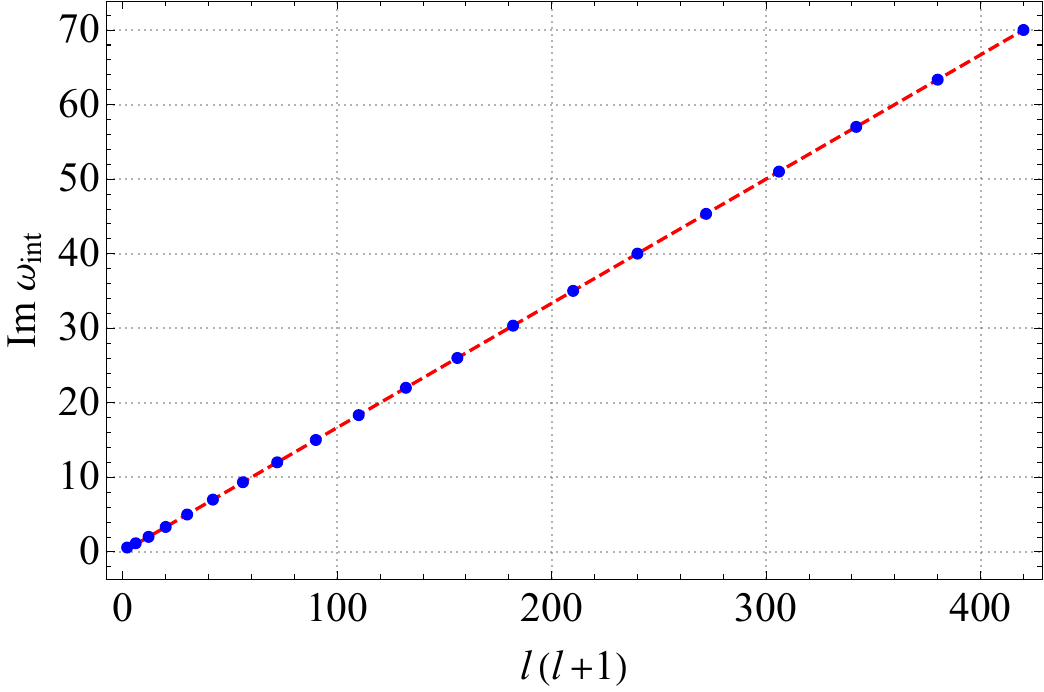}
\caption{Imaginary part of $\omega_{\rm int}$ for the axial fundamental mode as function of multipole number $l$, using the parameter setting $r_+=6$ and $\mu=1/10$. The red dashed line is the linear fit: $0.167\,l(l+1)+0.0203$.}
\label{fig:axial multipole}
\end{figure}

The diagnostics of instabilities afforded by the spectral analysis is corroborated by the solutions of the PDE in the time domain. As Fig.\ \ref{fig:axial time} shows, the function $u_A$ grows exponentially at late times, and slope measurements agree with the QNM spectrum, cf.\ Table \ref{tab:table2}. On the other hand, from Fig.\ \ref{fig:axial multipole} we see that the destabilization rate depends on the multipole number $l$. A fit of the numerical results shows a very clear linear dependence on $l(l+1)$, which is the combination that enters the effective potential, cf.\ Eq.\ \eqref{eq:axialequ}. The linear behavior is consistent with the analysis of dispersion relations in the short-wavelength approximation, which has the form $\omega^2\simeq k^2+\frac{l(l+1)}{r_+(r-r_+)}$ in the vicinity of the pole \cite{Garcia-Saenz:2021uyv}. It follows that, for fixed $l$, perturbations localized near the pole interior experience the fastest instability rate. We may therefore estimate $(r-r_+)=-|r-r_+|\sim 1/k$ for the typical wavelength. Then the above expression for $\omega^2$ is minimized at $\omega_{\rm int}^2\sim -\left[\frac{l(l+1)}{r_+}\right]^2$, i.e.\ ${\rm Im}\,\omega_{\rm int}\propto l(l+1)$. We remark that unstable QNMs with instability rates that grow with multipole number have been previously studied in \cite{Konoplya:2018qov}.

We can actually improve on the short-wavelength approximation through a more careful analysis of the mode equation, which allows us to explain the slope and intercept of the linear fit in Fig.\ \ref{fig:axial multipole}. Referring the reader to the Supplemental Material for the derivation, we have the analytical estimate $\omega_{\rm int}\simeq \left[\frac{l(l+1)}{r_+}+\frac{\mu^2r_+}{3}\right]i$. This expression is valid for large $l$. Comparison with the numerical fit shows perfect agreement for the slope ($\simeq 1/r_+$), which moreover matches the back-of-the-envelope estimate of the previous paragraph. The prediction for the intercept also agrees reasonably well with the numerical results.

\begin{table}
\begin{center}
\begin{tabular}{|c|c|c|}
\hline
$l=1$, $\mu=0$, $r_+=5$ & $\omega_{\rm int}$ & $\omega_{\rm ext}$ \\ \hline
Spectral method &  $0.6091i$ & $\pm 0.02214-0.2181i$ \\ \hline
Prony method &  $0.6061i$ & $\pm 0.02451-0.2004i$ \\ \hline \hline
$l=1$, $\mu=1/10$, $r_+=5$ & $\omega_{\rm int}$ & $\omega_{\rm ext}$ \\ \hline
Spectral method & $0.6098 i$ & $\pm0.01657-0.2099 i$ \\ \hline
Prony method & $0.6066 i$ &  $?-0.2107 i$ \\ \hline
\end{tabular}
\end{center}
\caption{Comparison of results for the fundamental QNM frequency of axial-vector perturbations, using particular parameter settings, for the spectral and Prony methods. The question mark indicates that the Prony method fails to extract the real part of the frequency, as discussed in the main text.}
\label{tab:table2}
\end{table} 

We find no qualitative novelties in the exterior region $r\in(r_+,\infty)$ relative to the results for the monopole mode. In agreement with expectations, the spectrum is stable, with a dependence of the QNM frequency $\omega_{\rm out}$ on the pole location. From the time domain evolution we observe again the interplay of oscillations modulated by $\omega_{\rm out}$ and the mass scale $\mu$. Unlike the monopole, the axial mode has a consistent massless limit, which manifests itself in the time domain through the absence of the latter $\mu$ tail; see the Supplemental Material. The application of the Prony method confirms that the QNM spectrum serves as an accurate measure of the behavior of perturbations under time evolution, as shown by the agreement of the two calculations, cf.\ Table \ref{tab:table2}. However, we find that the precision of the method decreases as the Proca mass increases, as the QNM behavior is quickly hindered by the $\mu$ tail.

%-------------------------------
%-------------------------------

\section{Discussion}

We have investigated the phenomenon of trapped QNMs as a novel destabilization mechanism of black holes. Although the focus of this work was on the non-minimally coupled Einstein-Proca theory, we naturally expect trapped QNMs to exist quite generically in models characterized by non-minimal gravitational couplings. In fact, it is not hard to devise toy models that fall into this category. For the sake of illustration, in the Supplemental Material we briefly discuss a simple scalar-tensor theory that accommodates trapped QNMs in a Schwarzschild background. Interestingly, this set-up features some qualitative differences relative to the vector-tensor one: we find an instability for all values of the pole distance, similarly to the axial-vector case and in contrast to the monopole one, yet the destabilization rate is controlled by the mass of the field rather than by the multipole number. We plan to delve into this property in a forthcoming work \cite{inprep1}, where the reader may also find further results as well as extensions of the present analysis. In particular, there we elaborate and improve on our analytical formulae.

Our work may be generalized in several directions. The inclusion of a cosmological constant would be straightforward. In the case of anti-de Sitter space, this could find applications in holographic condensed matter systems \cite{Hartnoll:2009sz,Hartnoll:2016apf}. We have also already mentioned the case of de Sitter space as a interesting variation on the theme of trapped QNMs because of the presence of a second horizon. In relation to this, we remark that the polar-vector mode equations are sensitive to both poles $r_{\pm}$ \cite{Garcia-Saenz:2022wsl}. However, since both cannot lie in the physical domain, we do not expect novel features in comparison with the axial modes. On the other hand, in systems with several non-minimal couplings, and thus several scales, one may foresee the possibility of having two or more poles outside the event horizon, giving rise to a new type of boundary value problem in the region between two such poles.

On a more physical note, it would be critical to understand how trapped QNMs evolve beyond the linearized regime, in particular whether the instability is potentially quenched by non-linear effects or if it should instead be seen as catastrophic, hence ruling out or at least constraining a wide class of models of modified gravity and cosmology. This could be investigated by taking into account the backreaction of the field on the spacetime geometry in a perturbative fashion, or more ambitiously by full numerical relativity calculations \cite{East:2017ovw}. A related question concerns the effects of trapped QNMs from the perspective of an observer in the exterior region. Although the vector perturbation is indeed trapped (at linear order), it could still couple with other matter fields, whose equations are not subject to extra poles and are therefore free to propagate to spatial infinity. In turn, these matter fields will themselves source gravitational waves outside the pole, which will therefore also indirectly carry imprints of the instability. It would be interesting to achieve a precise characterization of such observables in future work.

%======================
%======================

\begin{acknowledgments}
The authors would like to thank Jun Zhang for some very helpful conversations and comments. The work of HWC, SGS and AS is supported by the NSFC (Grant No.\ 12250410250). SGS also acknowledges support from a Provincial Grant (Grant No.\ 2023QN10X389).
\end{acknowledgments}

\appendix

\section{Outside the pole}

\begin{figure}[t]
	\centering
	\includegraphics[width=0.48\textwidth]{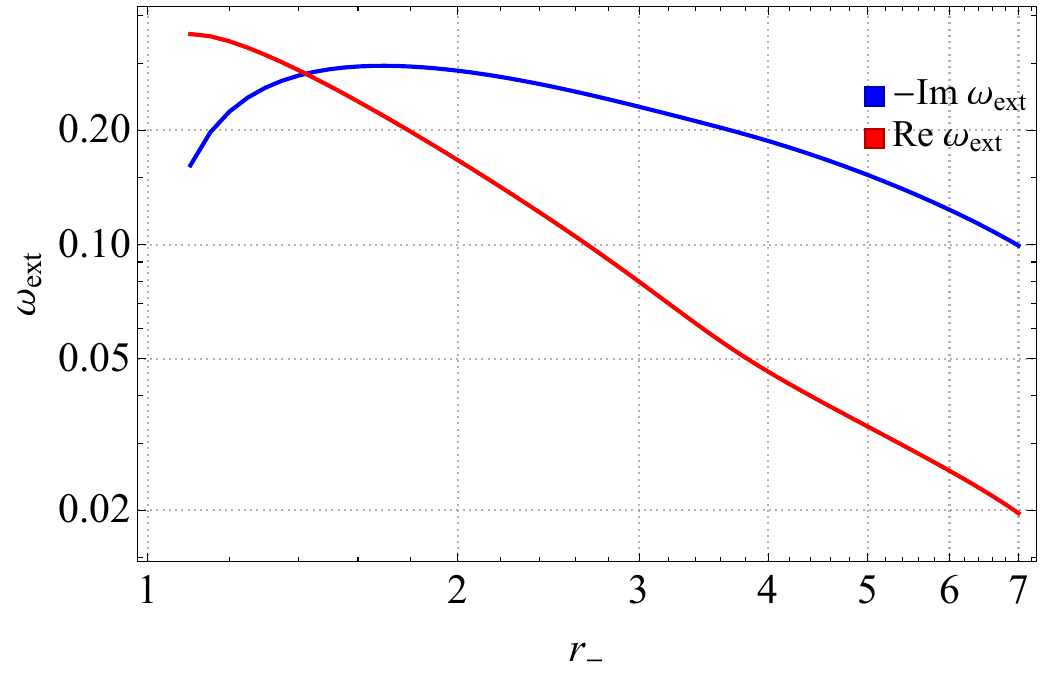}
	\caption{Dependence of the ``exterior'' frequency $\omega_{\rm ext}$ of monopole perturbations on the value of $r_-$, choosing $\mu=1/10$ for the Proca mass. The curves show the imaginary (blue) and real (red) parts of the fundamental mode.}
	\label{fig:monopoleoutside}
\end{figure}

\begin{figure}[t]
	\centering
	\includegraphics[width=0.48\textwidth]{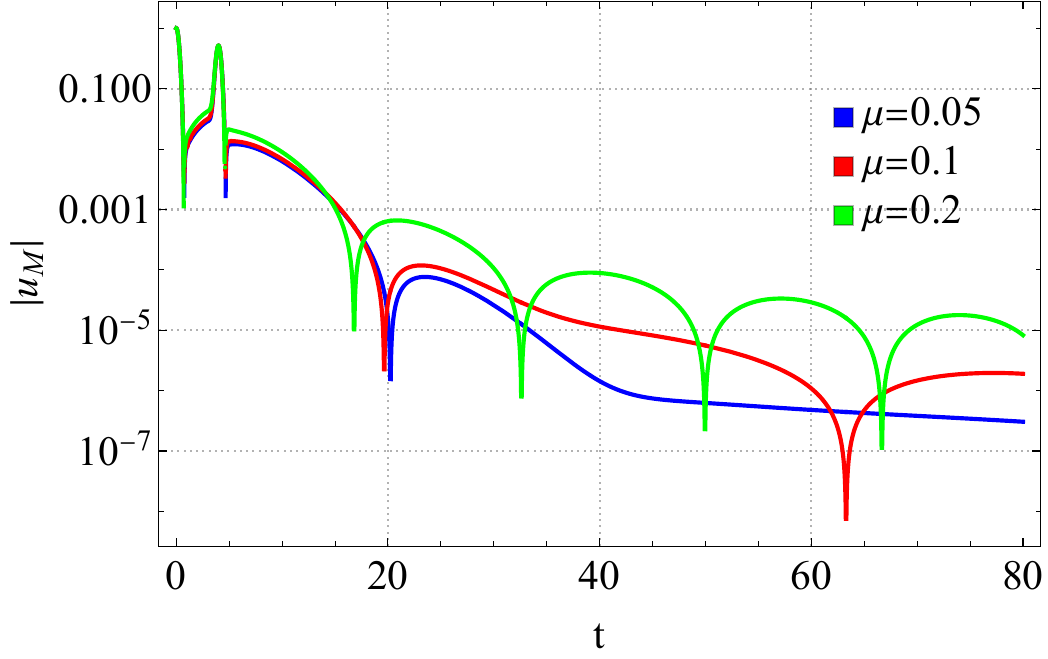}
	\caption{Time evolution of $u_M$ outside the pole. The parameter setting is $r_-=2$, $a=2$ and $\Delta=1/3$, with varying values of $\mu$. The mode function is evaluated $r_\ast=a$.}
	\label{fig:mono-2}
\end{figure}

\begin{figure}[t]
	\centering
	\includegraphics[width=0.47\textwidth]{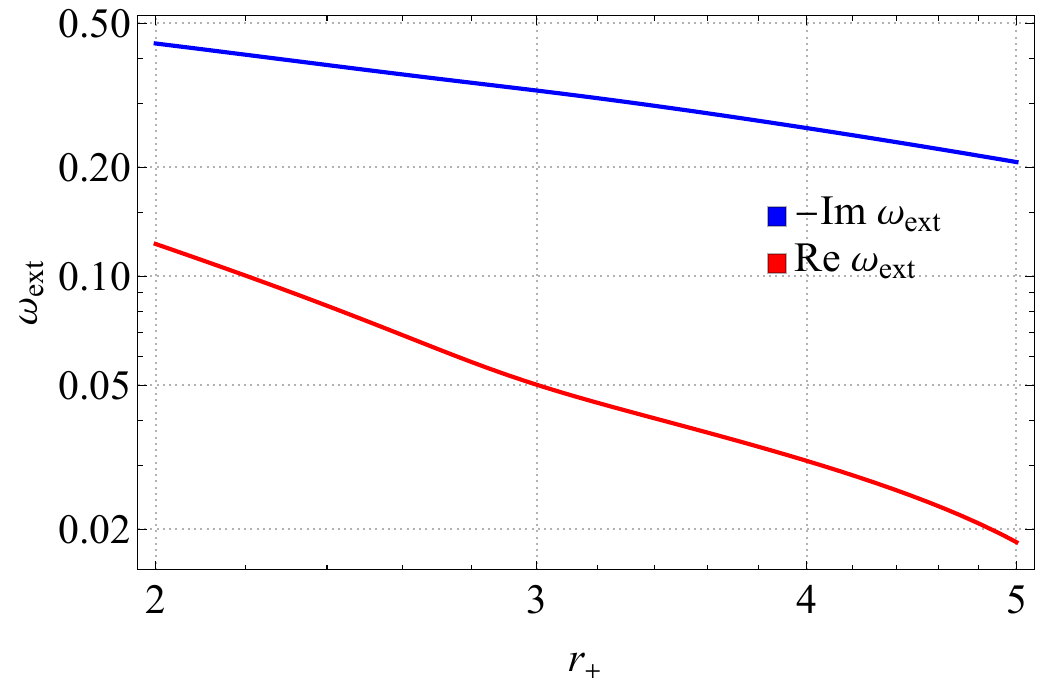}
	\caption{Dependence of the ``exterior'' frequency $\omega_{\rm ext}$ of axial-vector perturbations on the value of $r_+$, choosing $\mu=1/10$ and $l=1$. The curves show the imaginary (blue) and real (red) parts of the fundamental mode.}
	\label{fig:axial omega out}
\end{figure}

\begin{figure}[t]
	\centering
	\includegraphics[width=0.48\textwidth]{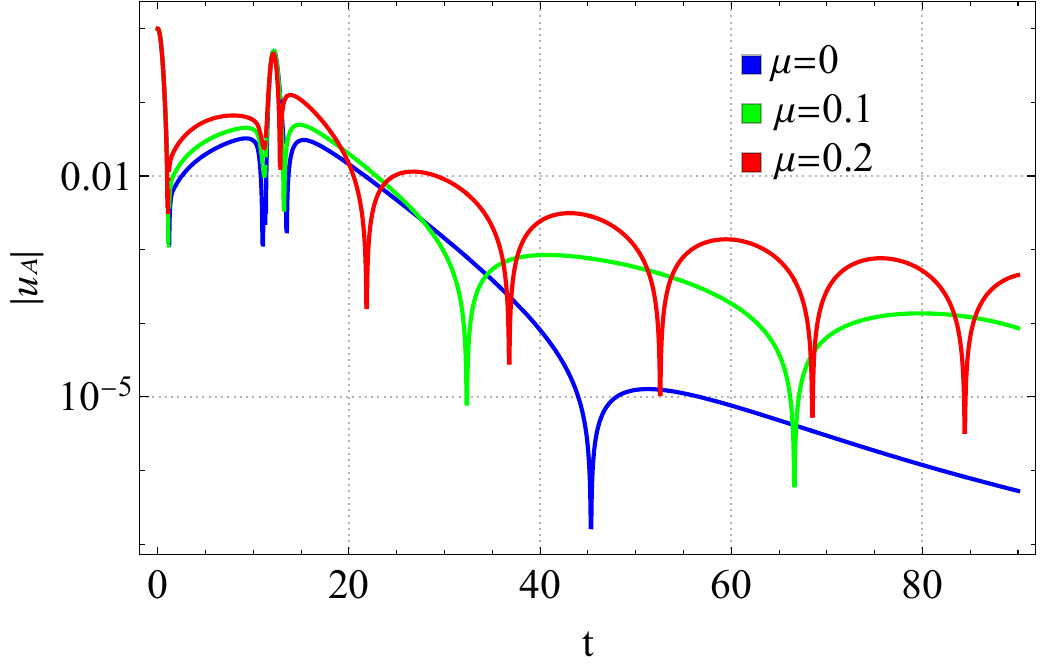}
	\caption{Time evolution of $u_A$ outside the pole. The parameter setting is $r_+=5$, $a=6$ and $\Delta=1/2$, with varying values of $\mu$. The mode function is evaluated $r_\ast=a$.}
	\label{fig:axial time out}
\end{figure}

For the sake of brevity, we chose to omit in the main text the plots for the results of our analysis in the pole exterior. In Fig.\ \ref{fig:monopoleoutside} we display the fundamental mode frequency of monopole perturbations as function of the pole distance $r_-$, while Fig.\ \ref{fig:mono-2} shows the results of the integration in the time domain for the same mode using different parameter settings. Figs.\ \ref{fig:axial omega out} and \ref{fig:axial time out} correspond to the same plots but for the axial mode outside the pole.

\section{Numerical calculations} \label{sec:numerics}

In this Appendix we provide details on our numerical schemes and set-up, as well as information on the convergence of the spectral method and error estimates of the time-domain method.

\subsection{Spectral method}

QNM frequencies are determined by a spectral method based on Chebyshev interpolation; see \cite{boyd,trefethen} for textbook introductions.

We study a generic boundary value problem for an equation of the form
\be \label{eq:spectral ode}
{\psi}''(x)+p(\omega,x){\psi}'(x)+q(\omega,x){\psi}(x)=0 \,,
\ee
where $x\in[-1,1]$, $\psi$ is assumed analytic in this domain, and $\omega$ is a complex parameter one seeks to determine. The domain is discretized into a Chebyshev grid defined by
\be
x_n=\cos\left(\frac{n}{N}\pi\right) \,,\qquad n=0,1,2,\ldots,N \,,
\ee
and the function $\psi(x)$ admits an approximate representation given by
\be \label{eq:spectral discretization}
\psi_N(x)\equiv \sum_{n=0}^N \psi(x_n)C_n(x)\,,
\ee
where the cardinal functions are defined as
\be
C_n(x)\equiv \cos\left(n\arccos(x)\right) \,.
\ee
It may be shown that $\psi_N$ converges uniformly to $\psi$ in the limit $N\to\infty$ \cite{boyd}.

Substituting \eqref{eq:spectral discretization} in \eqref{eq:spectral ode} results in a set of algebraic equations,
\be \label{eq:spectral algebraic}
\sum_{n=0}^N M_{mn}(\omega){\psi}(x_n)=0\,,    
\ee
where the elements $M_{mn}$ are determined by
\be
M_{mn}(\omega)= C''_n(x_m)+p(\omega,x_m)C'_n(x_m)+q(\omega,x_m)C_n(x_m) \,.
\ee
Non-trivial solutions of \eqref{eq:spectral algebraic} are given by $\det M(\omega)=0$, which determines the spectrum $\{\omega\}$.

%------------------------------

\subsection{Time-domain method}

To solve the mode equations in the time domain we utilize a finite difference method in the radial direction and the default ODE integration method of Mathematica (Adams) along the time direction. The radial grid is chosen to have uniform spacing in the tortoise coordinate $r_\ast$, starting at a narrow distance from the pole, and is wide enough so that the two endpoints remain causally disconnected throughout the evolution. Without loss of generality, we set $r_\ast = 0$ at the pole $r_\pm$.

As explained in the main text, the standard choice of initial data is given by a Gaussian waveform with zero velocity,
\be \label{eq:initialcondition}
u (0,r_\ast) = \exp\left[-\left(\frac{r_\ast - a}{\Delta}\right)^2\right] \,,\quad \dot{u} (0,r_\ast) = 0  \,.
\ee
This choice leads to two initial superposed modes, one moving outward and one inward. Time evolution then eventually results in five components: two initial Gaussian waves, two associated late-time tails (if they exist), and the QNM component.

\begin{figure}[t]
	\centering
	\includegraphics[width=0.48\textwidth]{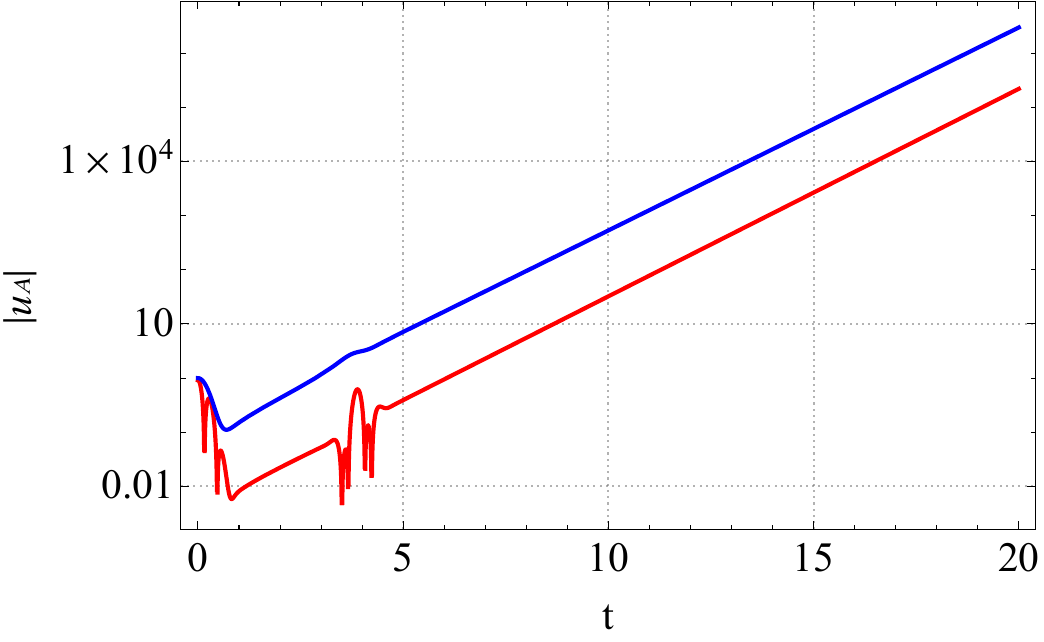}
	\caption{Time evolution of $u_A$ inside the pole. The parameter setting is $\mu=1/10$, $l=1$, $a=-2$, $\Delta=1/3$ and $r_+=2$. The mode function is evaluated $r_\ast=a$. The two curves correspond to different choices for the parameter $b$ in Eq.\ \eqref{eq:initialsin}: $b=0$ (blue) and $b=10$ (red).}
	\label{fig:app axials}
\end{figure}

Our calculations have been verified to be robust under changes in the initial data, i.e.\ the parameters $a$ and $\Delta$. Furthermore, we have verified some of the calculations with the more general initial profile
\be \label{eq:initialsin}
u(0,r_\ast)=\exp\left[-\left(\frac{r_\ast - a}{\Delta}\right)^2\right]\sin(b r_\ast) \,,
\ee
corresponding to a wave-packet that is shifted by $\pm b$ in momentum space. Despite very different evolutions on short time-scales, the behavior at late times becomes universal and consistent with the predictions of the spectral analysis. See Fig.\ \ref{fig:app axials} for an example.

%-----------------------------------

\subsection{Prony's method}

Prony's method allows one to extract the QNM frequencies from the time domain profile. We refer the reader to \cite{Berti:2007dg,Konoplya:2011qq} for modern expositions and references to the mathematical literature. In brief, the method consists of considering a waveform $\phi(t)$ in the discretized time interval $t\in\{t_0,t_0+h,\ldots,t_0+Nh\}$, where $N$ is an integer and $h$ is the step size. The waveform is then approximated by a finite sum of complex exponentials,
\be
\phi(t)\approx \phi_N(t)\equiv \sum_{n=1}^{p}c_n e^{- i \omega_n t} \,,
\ee
where $p\equiv\left\lfloor{N/2}\right\rfloor$. Define
\be
x_n\equiv \phi(t_0+nh)=\sum_{k=1}^{p}\tilde{c}_k z_k^n \,,
\ee
where $\tilde{c}_k={c_k}e^{- i \omega t_0}$ and $z_k=e^{i \omega_k h}$, and let
\be \label{eq:prony A}
A(z)=\prod_{n=1}^p(z-z_n)=\sum_{n=0}^p \alpha_n z^n \,,
\ee
which defines also the sequence $\{\alpha_n\}_{n=0}^p$. Noting that
\be
\sum_{n=0}^{p}\alpha_n x_{n+m}=\sum_{k=1}^{p}\tilde{c}_k z_k^m A(z_k)=0 \,,
\ee
and $\alpha_p=1$, we obtain $p$ equations for the unknowns $\{\alpha_n\}_{n=0}^{p-1}$. After solving these, Eq.\ \eqref{eq:prony A} then determines the parameters $z_k$, and hence the frequencies $\omega_k$.

%-------------------------------------

\subsection{Convergence tests for spectral method}

We comment here on the convergence of our spectral method calculations. The test that a result must pass for it to be accurate is that the quantity $|\omega_{{\rm Max}\,N}-\omega_N|$ must decrease as $N$ increases, or at least not increase or oscillate randomly. Here $N$ denotes the number of nodes in the Chebyshev grid, $\omega_N$ is the frequency computed with this grid size, while $\omega_{{\rm Max}\,N}$ is the frequency computed with some large value ${\rm Max}\,N$; see e.g.\ \cite{Baumann:2019eav}.

\begin{figure}[t]
	\centering
	\includegraphics[width=0.46\textwidth]{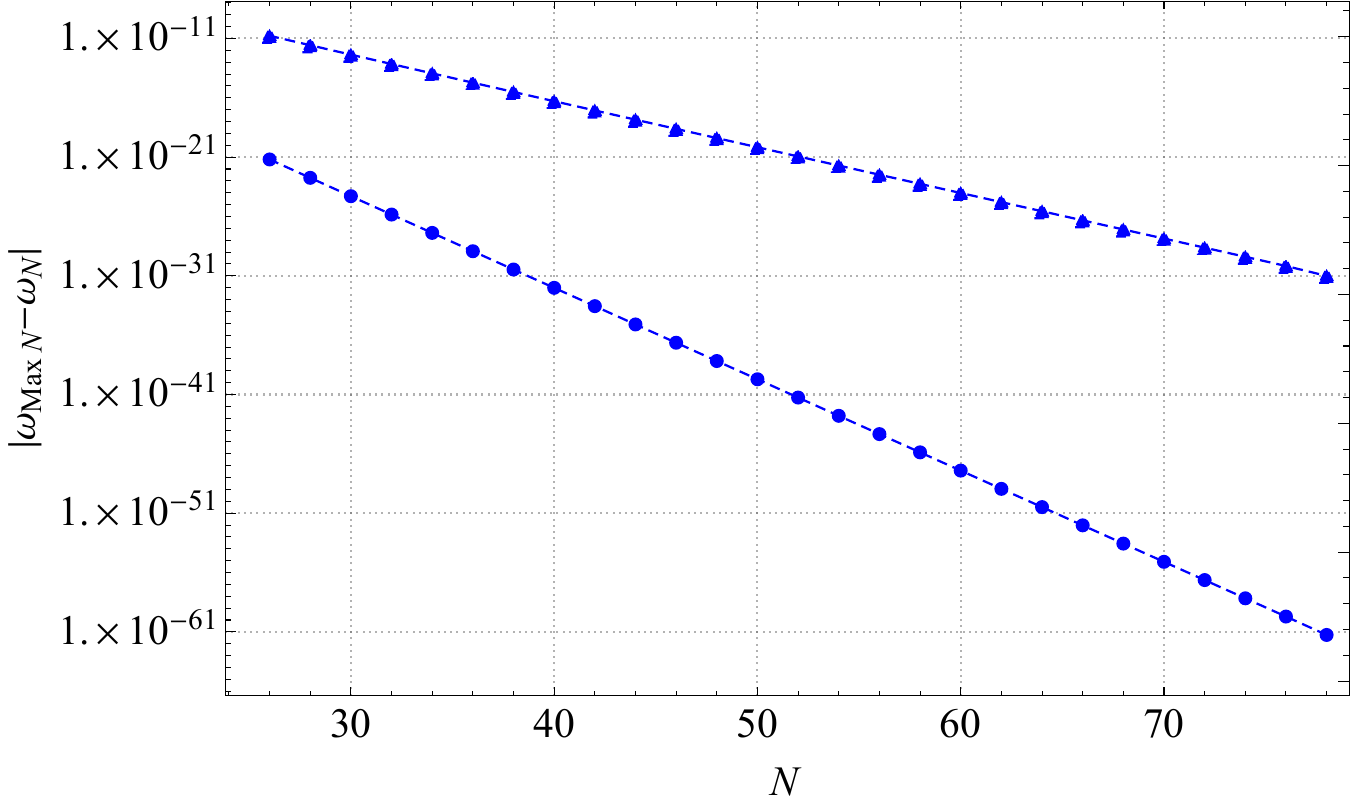}
	\caption{Convergence test for monopole QNM frequency of fundamental mode in the pole interior, using $\mu=1/10$ and ${\rm Max}\,N=80$. The curves show the imaginary part for $r_-=2$ (circles) and $r_-=6$ (triangles).}
	\label{fig:test mono int}
\end{figure}

\begin{figure}[t]
	\centering
	\includegraphics[width=0.46\textwidth]{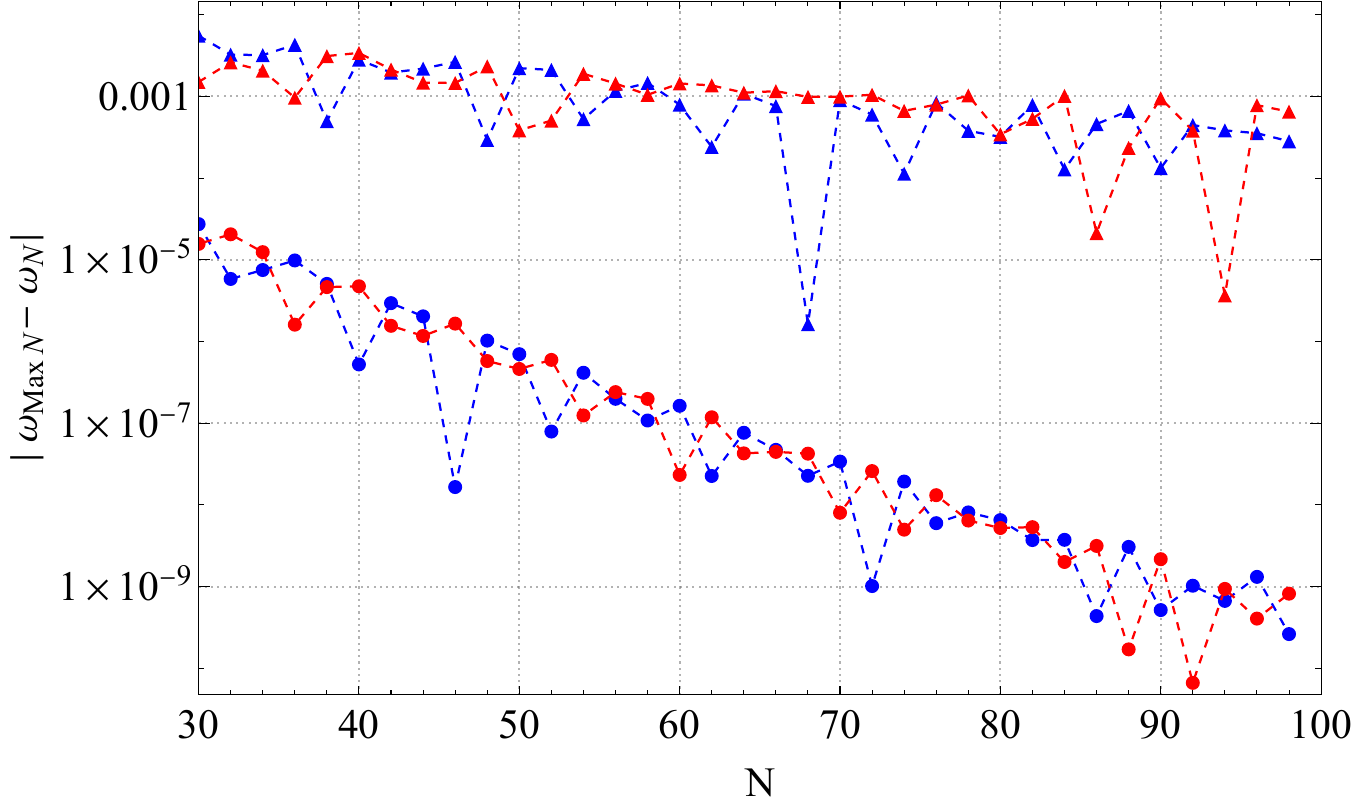}
	\caption{Convergence test for monopole QNM frequency of fundamental mode in the pole exterior, using $\mu=1/10$ and ${\rm Max}\,N=100$. The curves show the imaginary (blue) and real (red) parts for $r_-=2$ (circles) and $r_-=6$ (triangles).}
	\label{fig:test mono ext}
\end{figure}

\begin{figure}[t]
	\centering
	\includegraphics[width=0.46\textwidth]{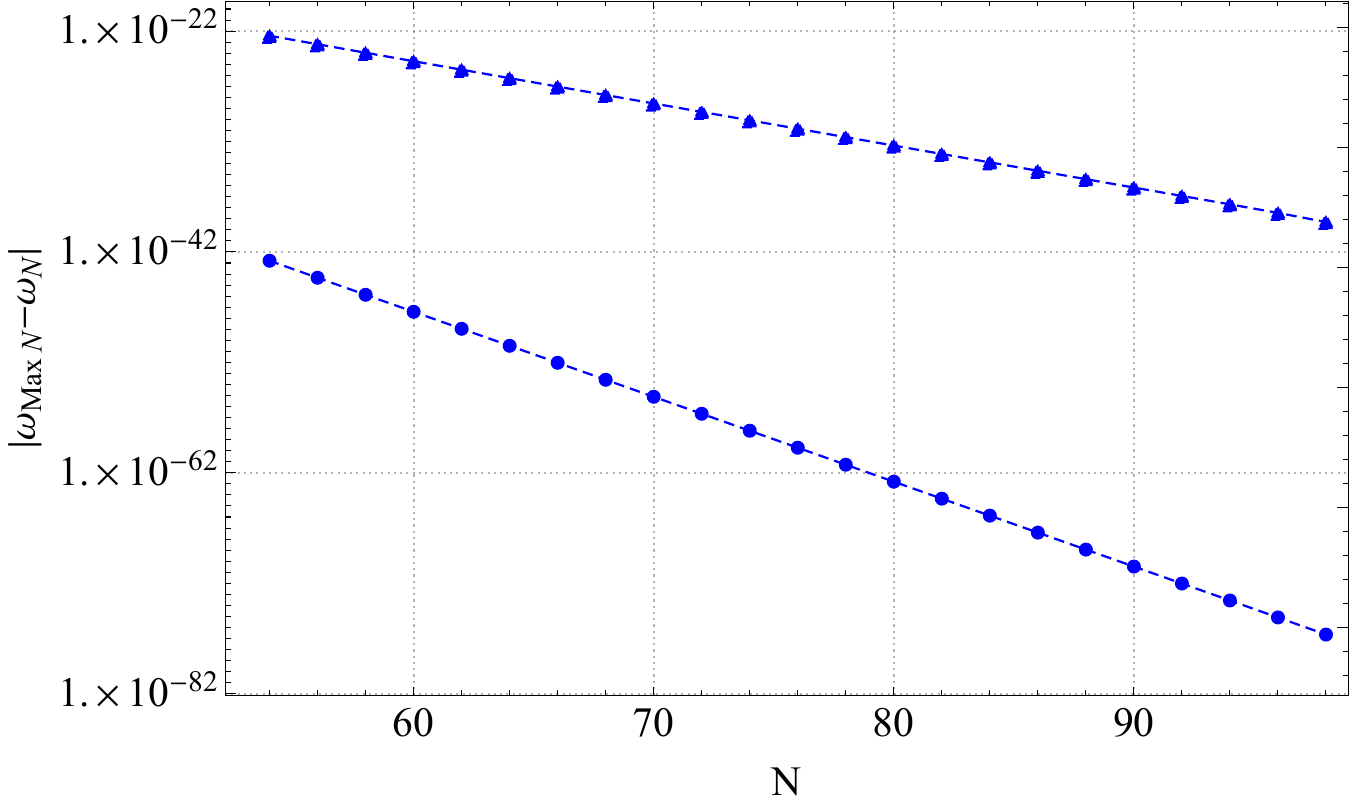}
	\caption{Convergence test for axial QNM frequency of fundamental mode in the pole interior, using $l=1$, $\mu=1/10$ and ${\rm Max}\,N=100$. The curves show the imaginary part for $r_-=2$ (circles) and $r_-=6$ (triangles).}
	\label{fig:test axial int}
\end{figure}

\begin{figure}[t]
	\centering
	\includegraphics[width=0.46\textwidth]{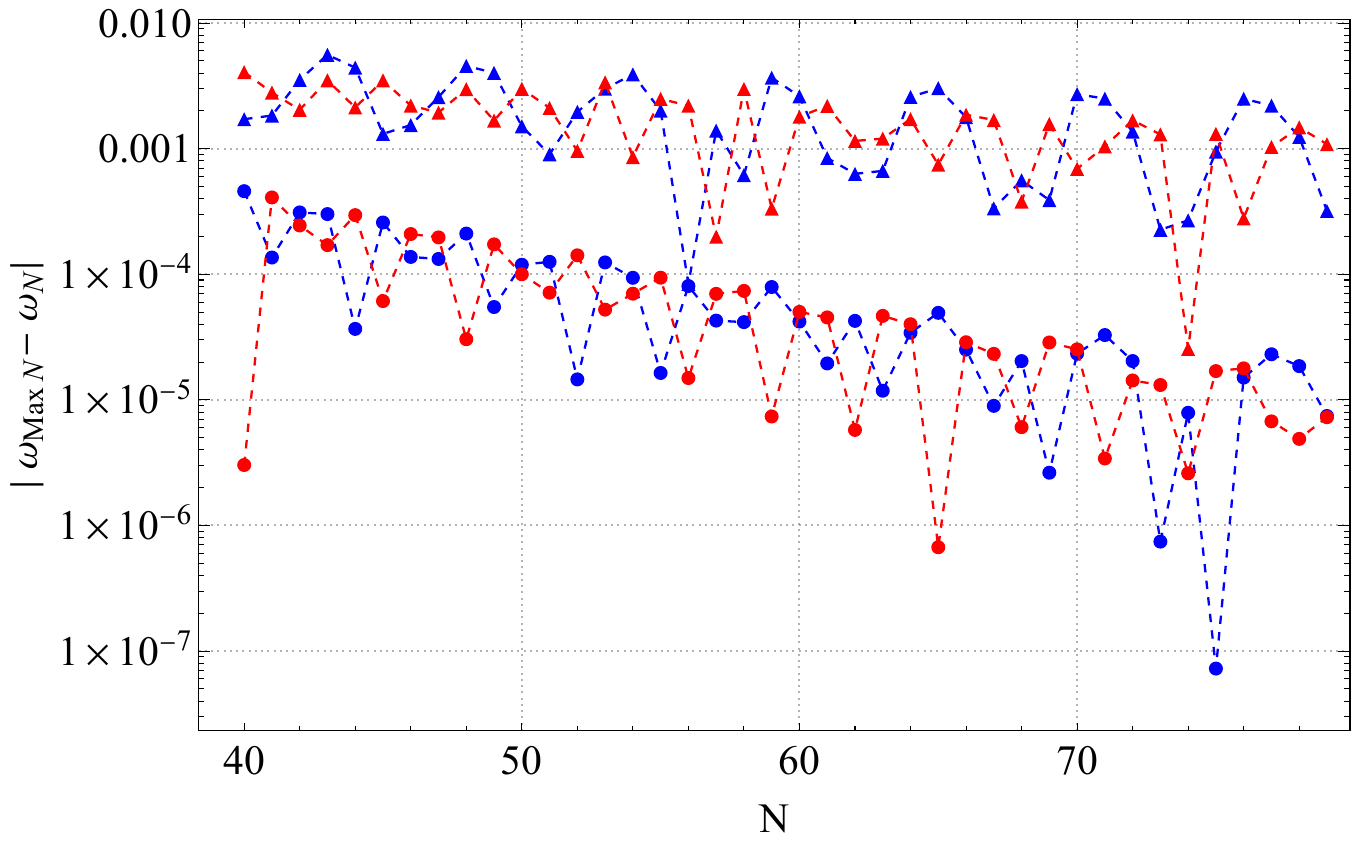}
	\caption{Convergence test for axial QNM frequency of fundamental mode in the pole exterior, using $l=1$, $\mu=1/10$ and ${\rm Max}\,N=80$. The curves show the imaginary (blue) and real (red) parts for $r_-=2$ (circles) and $r_-=6$ (triangles).}
	\label{fig:test axial ext}
\end{figure}

Spectral analysis in the pole interior shows very good convergence, as is apparent from Figs.\ \ref{fig:test mono int} and \ref{fig:test axial int}. Convergence in the pole exterior, cf.\ Figs.\ \ref{fig:test mono ext} and \ref{fig:test axial ext}, is on the other hand much slower. This is expected from the fact that the norm of the frequency becomes small as the pole distance increases, as seen in Figs.\ \ref{fig:monopoleoutside} and \ref{fig:axial omega out}. In spite of this, we still find that the error remains under control and displays a clear decreasing trend, provided the pole radius is chosen not too large.

%--------------------------------------

\subsection{Time-domain error analysis}

In this section we discuss the precision and accuracy of our results for the time-domain integration and Prony methods. We comment on (i) the error due to the discretization in the radial direction, (ii) the effect of our regularization scheme at the pole, and (iii) the finite window effect on the QNM extraction with the Prony method.

\subsubsection*{Discretization error}

The fourth order finite difference operator for the differential equation $-\partial_t^2 u (t,r_\ast) + \partial_{r_\ast}^2 u (t,r_\ast) - V(r_\ast) u (t,r_\ast) = 0$ is given by
\be\begin{aligned} \label{eq:finite difference delta}
	\;&\Delta_h u (t, r_\ast) = \frac{1}{12 h^2} \Big[ - u(t, r_\ast - 2h) + 16 u(t, r_\ast - h) \\
	&- 30 u(t, r_\ast) + 16 u(t, r_\ast + h) - u(t, r_\ast + 2h) \Big] \,,
\end{aligned}\ee
where $h$ is the grid spacing size. The PDE is then discretized into a set of ODEs: $-\partial_t^2 u_I (t) + \sum_{J}\Delta_{\epsilon, IJ} u_J (t) - V_I u_I(t) = 0$, where $I$-th mode $u_I$ corresponds to the mode function at $r_\ast = \epsilon + h I$, and $V_I$ is the potential evaluated at this radius. Here $\epsilon$ denotes the distance from the edge of the grid to the pole (see below for more details on this) and $\Delta_{\epsilon, IJ}$ is the matrix of coefficients of the finite difference operator \eqref{eq:finite difference delta}.

\begin{figure}
	\centering
	\includegraphics[width=0.47\textwidth]{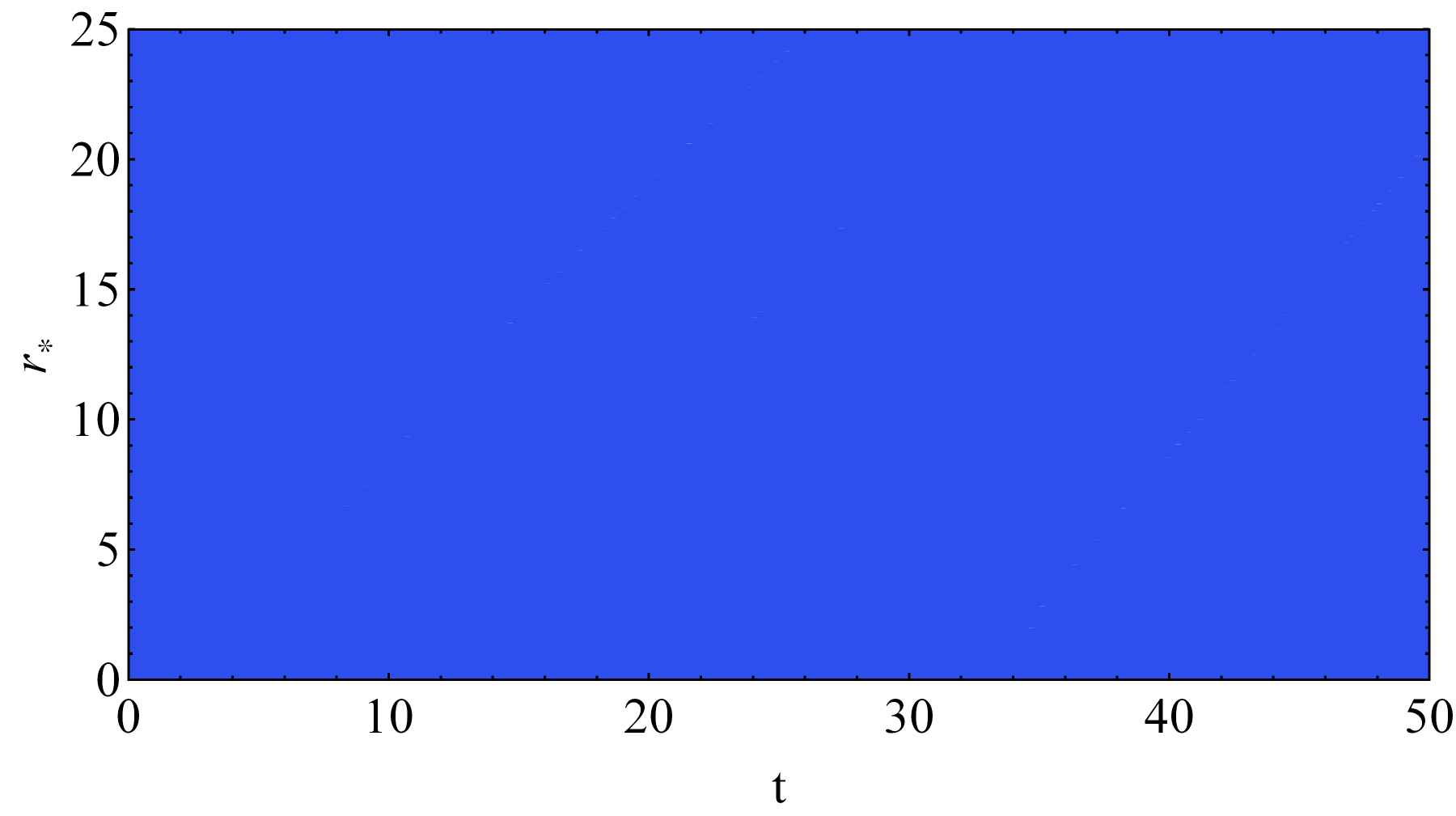}\\
	~~~\includegraphics[width=0.455\textwidth]{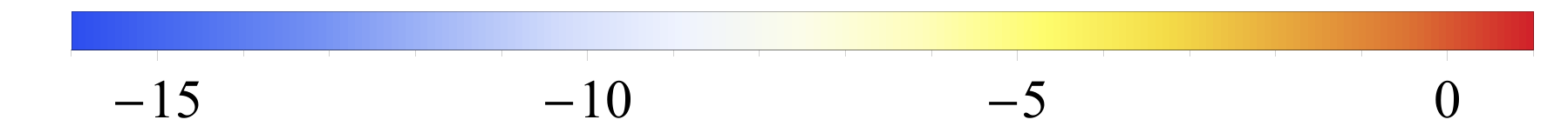}
	\caption{Discretization error $\log_{10}\mathcal{E}_h$ for the axial mode equation in the pole exterior, with the parameter setting $l = 1$, $\mu = 1/10$, $r_+ = 5$, $\Delta=1/2$, $a = 6$, $h = 1/80$ and $\epsilon = 10^{-8}$.}
	\label{fig:error_spatial}
\end{figure}

The discretization error may be defined as the residue of the differential equation resulting from the discretization process \cite{BANKS20121}:
\begin{align} \label{eq:discretization error}
	\mathcal{E}_h \equiv \frac{\left| \Delta_h u - \partial_{r_\ast}^2 u \right|}{{\rm ULP}(u) + \left| \Delta_h u \right|} = \frac{\left| \Delta_h u - \partial_t^2 u - Vu \right|}{{\rm ULP}(u) + \left| \Delta_h u \right|} \,,
\end{align}
where ULP stands for ``unit at the last place'', equal to $2^{-52} \sim 10^{-16}$ for the double precision ODEPACK library of Mathematica \cite{osti_145724}. Notice that the numerator of $\mathcal{E}_h$ is precisely the ODE set after discretization. As we apply the Adams solver in the time domain \cite{osti_4384150}, $\mathcal{E}_h$ can be well estimated through spline interpolation in the same basis \cite{SHAMPINE197547}. In Fig.~\ref{fig:error_spatial} we show the discretization error for the axial mode equation in the pole exterior. It is clearly seen that the error is $\mathcal{O}({\rm ULP})$ everywhere, except when $u$ approaches zero, which results in $\mathcal{E}_h \sim \sqrt{\rm ULP} \sim 10^{-8}$ (not visible in the figure as it occurs on very localized regions in the $t-r_\ast$ plane), as expected for a second-order equation. Similar results have been obtained for the other equations considered in this paper.

\subsubsection*{Pole regularization}

In numerical calculations, boundary conditions cannot be set at the exact location of a pole. One considers instead a small displacement $\epsilon$, so that in the situation of Dirichlet boundary conditions that we use in this work, we have $u(t,r_\ast=\pm\epsilon)=0$ at the pole (where the sign depends on whether we consider the exterior or interior domain, and recall that we define the tortoise coordinate such that $r_\ast=0$ corresponds to the pole). In our calculations $\epsilon$ is chosen as $10^{-8}$, and we have verified the robustness of the results under changes in $\epsilon$, provided it is not much larger than this value. Furthermore, the boundary condition remains tightly satisfied over a long enough period, ensuring the soundness of our time-domain analysis.

Formula \eqref{eq:discretization error} for the discretization error does not apply very close to the pole. In this case, a good measure of the precision of the numerics is provided by the logarithmic gradient, $d\log u/d\log r_\ast$. From the analytical form of the near-pole solution, we may predict the limiting value of this quantity as one approaches the pole. We then compare this with the results of our numerical calculations with varying $\epsilon$, and assess whether it approaches the predictions as $\epsilon$ is made smaller. We find our results to be numerically consistent.

\subsubsection*{QNM extraction}

As we have seen, in the pole exterior region the norm of the QNM frequency becomes small for large values of the pole distance. This affects not only the precision of the spectral method calculation but also the QNM extraction via the Prony method. The reason is that the smallness of $|\omega|$ causes the waveform to become quickly dominated by the late-time tail associated to the Proca mass $\mu$, as explained in the main text. This should serve as word of caution regarding the precision of our calculations for the QNM frequencies with the Prony method. This issue appears to be less important for the monopole mode (cf.\ Table \ref{tab:table1}) and for the axial mode in the regime of very small $\mu$ (cf.\ Table \ref{tab:table2}).

We extract the $\mu$ tail by performing a fit of the form $u_{\rm tail}\propto t^{-q}e^{i\mu t}$ on a time window that is late enough so that the tail shows clear dominance. To a good approximation, our results are consistent with the prediction $q=l+1$ \cite{Leaver:1986gd}, which is moreover robust under changes in the fitting window. Note however that one does not expect an exact agreement, since it is known that the tail is in general sensitive to initial conditions \cite{Chavda:2024awq}.

%------------------------------------
%------------------------------------

\section{Scalar-tensor toy model} \label{sec:scalar toy model}

Although the focus of this paper was on the non-minimally coupled Einstein-Proca theory, we have emphasized that trapped QNMs are a very general phenomenon. As an interesting illustration, we consider a scalar-tensor theory described by the action
\begin{equation} \label{eq:scalar action}
	\begin{aligned}
		S[g,\phi]&=\int d^4x\sqrt{-g}\bigg[\frac{M_{\rm Pl}^2}{2}R-\frac{1}{2}\nabla^{\mu}\phi\nabla_{\mu}\phi-\frac{\mu^2}{2}\phi^2 \\
		&\quad +\gamma R^{\mu\rho\sigma\tau}R^{\nu}{}_{\rho\sigma\tau}\nabla_{\mu}\phi\nabla_{\nu}\phi\bigg] \,.
	\end{aligned}
\end{equation}
The derivation of the mode equation results in the following effective potential:
\be\begin{aligned}
	V= \frac{f}{r^2} \left[\frac{\left(l(l+1)+6\right) r-8}{r} + \frac{\mu^2r^2+3}{P}-\frac{9f}{P^2}\right] \,,
\end{aligned}\ee
for the mode function $u$ defined via $\phi = r^{-1}P(r)^{-1/2} \sum_{l,m} u(t, r) Y^{lm}$. Here we defined 
\be\begin{aligned}
	P(r)\equiv 1-\frac{r_p^6}{r^6} \,,\qquad r_p^6\equiv 6\gamma  \,.
\end{aligned}\ee
This model therefore exhibits a second-order pole at $r=r_p$. Assuming $\gamma>1/6$, the pole lies in the physical domain and we expect interesting effects associated to trapped QNMs in the region $r\in(1,r_p)$.

Before discussing the numerical calculations, it is interesting to comment on the predictions afforded by the short wavelength approximation. In this regime, we find the following result for the dispersion relation:
\be \label{eq:dispersion relation scalar}
\omega^2=f^2k^2+\frac{f}{r^2}\left[l(l+1)+\frac{\mu^2}{r^6-r_p^6}\right] \,,
\ee
where $k$ is the radial momentum. We thus see that the scalar mode should be free from pathologies associated to ghost or gradient instabilities. On the other hand, we note that the mass term has the wrong sign in the pole interior, suggesting the existence of unstable QNMs. We also remark that the gradient terms in the dispersion relation are independent of the non-minimal coupling constant. This may be understood from the fact that we can write the kinetic operators in \eqref{eq:scalar action} in terms of an effective metric $g_{\mu\nu}+2\gamma R_{\mu\rho\sigma\tau}R_{\nu}{}^{\rho\sigma\tau}$, which on the Schwarzschild background becomes conformally related to the Schwarzschild metric. This observation highlights that the existence of a pole in the mode equation does not necessarily translate into the existence of gradient-unstable solutions.

\begin{figure}[h]
	\centering
	\includegraphics[width=0.47\textwidth]{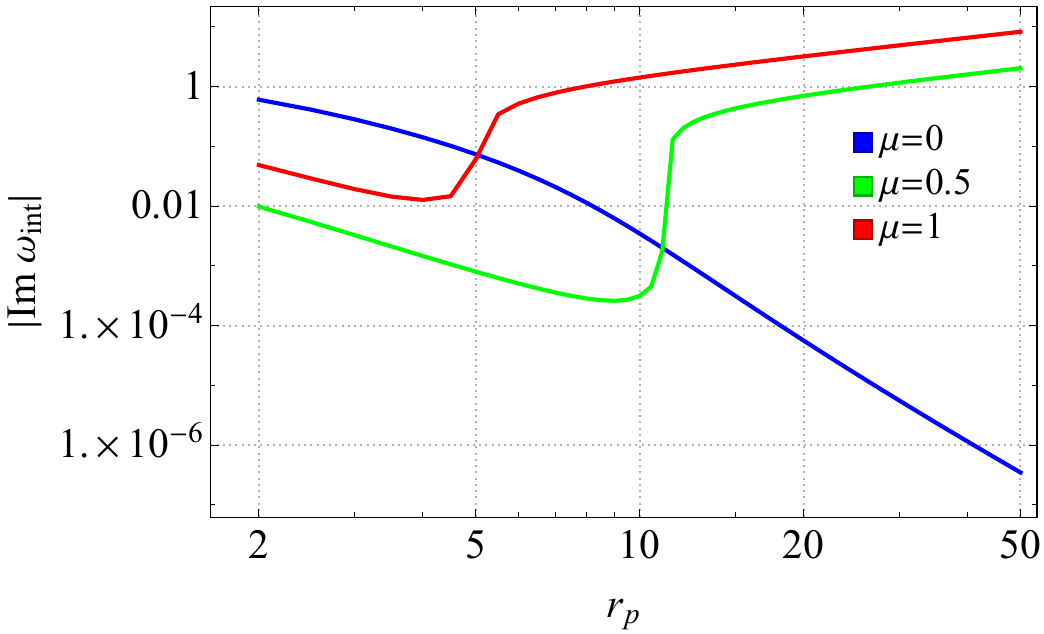}
	\caption{Dependence of the ``interior'' frequency $\omega_{\rm int}$ of the fundamental mode of scalar perturbations on the value of $r_p$, using $l=0$ and several values of $\mu$. ${\rm Im}\,\omega_{\rm int}$ is negative (stable) for $\mu=0$ and it is positive (unstable) for $\mu>0$.}
	\label{fig:scalar omega vs rp}
\end{figure}

\begin{figure}[h]
	\centering
	\includegraphics[width=0.47\textwidth]{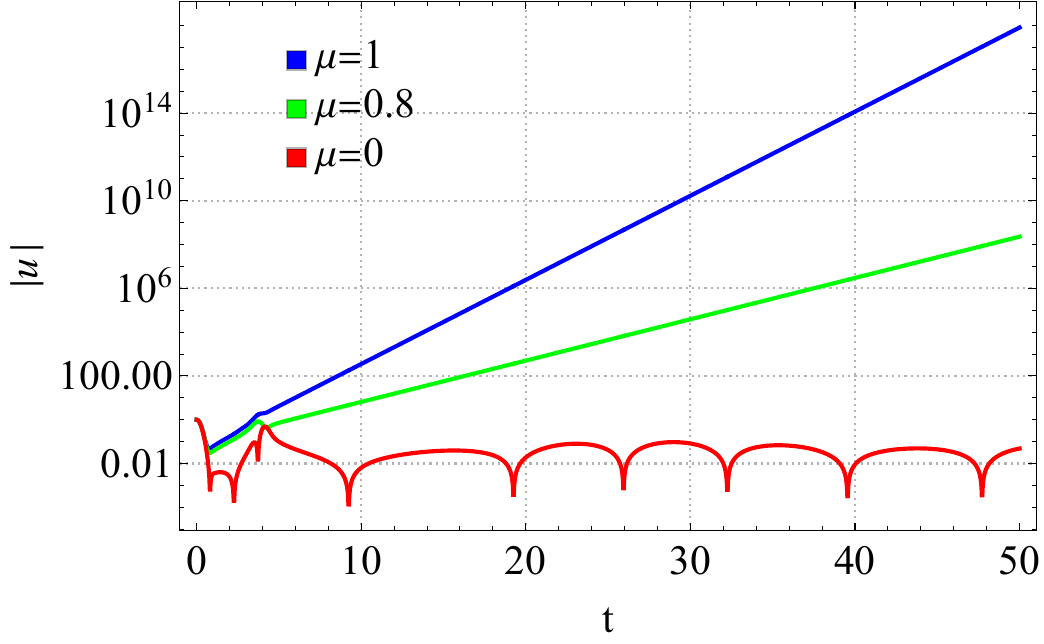}
	\caption{Time evolution of the scalar mode function inside the pole. The parameter setting is $l=0$, $a=-2$, $\Delta=1/3$ and $r_p=10$, using several values of $\mu$. The mode function is evaluated $r_\ast=a$.}
	\label{fig:scalar time domain}
\end{figure}

Turning to the numerical analysis in the frequency and time domains, we consider the boundary value problem discussed in the main text. For simplicity, here we focus only on the interior region, $r\in(1,r_p)$, and on the fundamental mode. Consistently with the intuition grasped from the dispersion relation \eqref{eq:dispersion relation scalar}, we find that trapped QNMs are unstable, with a growth rate that increases with the mass $\mu$. The exception is the massless case, $\mu=0$, which exhibits a stable spectrum; cf.\ Fig.\ \ref{fig:scalar omega vs rp}. Solutions of the mode PDE as functions of time confirm these results (see Fig.\ \ref{fig:scalar time domain}), once again highlighting the value of trapped QNMs as a diagnostic tool for instabilities.

%------------------------------------
%------------------------------------

\section{Near-pole expansion} \label{sec:near pole}

Consider a generic Schr\"{o}dinger-type equation $\frac{d^2u}{dr_\ast^2}+(\omega^2-V)u=0$, where we assume the potential admits the following expansion near $r_\ast=0$:
\be \label{eq:near pole potential}
V=\frac{V_{-2}}{r_\ast^2}+\frac{V_{-1}}{r_\ast}+V_0+\cdots \,.
\ee
Truncating the potential at the order shown results in a Whittaker equation, with solution
\be\begin{aligned}
	u&=c_1M_{-\frac{V_{-1}}{2\kappa},\frac{v}{2}}(2\kappa r_\ast)+c_2U_{-\frac{V_{-1}}{2\kappa},\frac{v}{2}}(2\kappa r_\ast) \,,
\end{aligned}\ee
where $M_{a,b}$ and $U_{a,b}$ are the Whittaker functions, $\kappa\equiv \sqrt{V_0-\omega^2}$ and $v\equiv \sqrt{1+4V_{-2}}\,$. Note that $M_{a,b}$ is regular at $r_\ast=0$ while $U_{a,b}$ is singular; hence we choose $c_2=0$ (and we set $c_1=1$ for simplicity). The function $M_{a,b}$ may be expressed in terms of the Laguerre function $L_\alpha^\beta$, so that
\be\begin{aligned}
	u&=\frac{\Gamma\left(1+v\right)\Gamma\left(\frac{1}{2}\left(1-v-\frac{V_{-1}}{\kappa}\right)\right)}{\Gamma\left(\frac{1}{2}\left(1+v-\frac{V_{-1}}{\kappa}\right)\right)} \\
	&\quad\times e^{-\kappa r_\ast}(2\kappa r_\ast)^{\frac{1}{2}\left(1+v\right)}L_{-\frac{1}{2}\left(1+v+\frac{V_{-1}}{\kappa}\right)}^{v}(2\kappa r_\ast) \,.
\end{aligned}\ee

We now make the assumption that the above solution is valid not only close to the pole but also in a certain domain where $\kappa r_\ast$ is large and negative. This is reasonable since we expect the solutions to be localized near the pole so that the structure of the potential far from it (i.e.\ the terms omitted in \eqref{eq:near pole potential}) should have a small impact on the form of the solution. Still, as we discuss below, this assumption will impose restrictions on the regime of applicability of the present analysis.

Expanding then at large $\kappa |r_\ast|$ (with $r_\ast<0$ in the pole interior), we get a combination of exponentially growing and decaying solutions,
\be\begin{aligned}
	u&\simeq -ie^{-i\frac{\pi}{2}\left(v+\frac{V_{-1}}{2}\right)}\frac{\Gamma\left(1+v\right)}{\Gamma\left(\frac{1}{2}\left(1+v-\frac{V_{-1}}{\kappa}\right)\right)}(2\kappa r_\ast)^{-\frac{V_{-1}}{2\kappa}}e^{-\kappa r_\ast} \\
	&\quad -e^{-i\frac{\pi}{2}\left(v-\frac{V_{-1}}{2}\right)}\frac{\Gamma\left(1+v\right)}{\Gamma\left(\frac{1}{2}\left(1+v+\frac{V_{-1}}{\kappa}\right)\right)}(2\kappa r_\ast)^{\frac{V_{-1}}{2\kappa}}e^{\kappa r_\ast} \,.
\end{aligned}\ee
Localized solutions in the potential ``well'' of the pole must decay as $r_\ast\to-\infty$, so we obtain the condition
\be
\frac{1}{2}\left(1+v-\frac{V_{-1}}{\kappa}\right)=-n \,,\qquad n=0,1,2,\ldots \,,
\ee
or, solving for $\omega^2$,
\be \label{eq:near pole general omega}
\omega^2=V_0-\frac{V_{-1}^2}{\left(2n+1+\sqrt{1+4V_{-2}}\right)^2} \,.
\ee

Naturally we do not expect this formula to provide a good approximation for the trapped QNM spectrum in every situation. First, we note that the validity of the near-pole expansion, Eq.\ \eqref{eq:near pole potential}, for large values of $\kappa |r_\ast|$ implies, in particular, that $\kappa^2 |V_{-2}|\gg\kappa |V_{-1}|\gg |V_0|$ (unless of course $V_{-2}=0$). Second, we have also assumed that the solution decays exponentially as one moves away from the pole. This requires that the frequency lie deep enough in the potential well so that the mode is always in the classically forbidden region away from the pole. In other words, if the potential has other wells \textit{besides} the pole (and let $V_{\rm min}$ be the minimum of all these wells, i.e.\ the minimum of $V$ \textit{without} the pole), then we must require $\omega^2<V_{\rm min}^2$ for the analysis to be self-consistent.

\begin{figure}
	\centering
	\includegraphics[width=\linewidth]{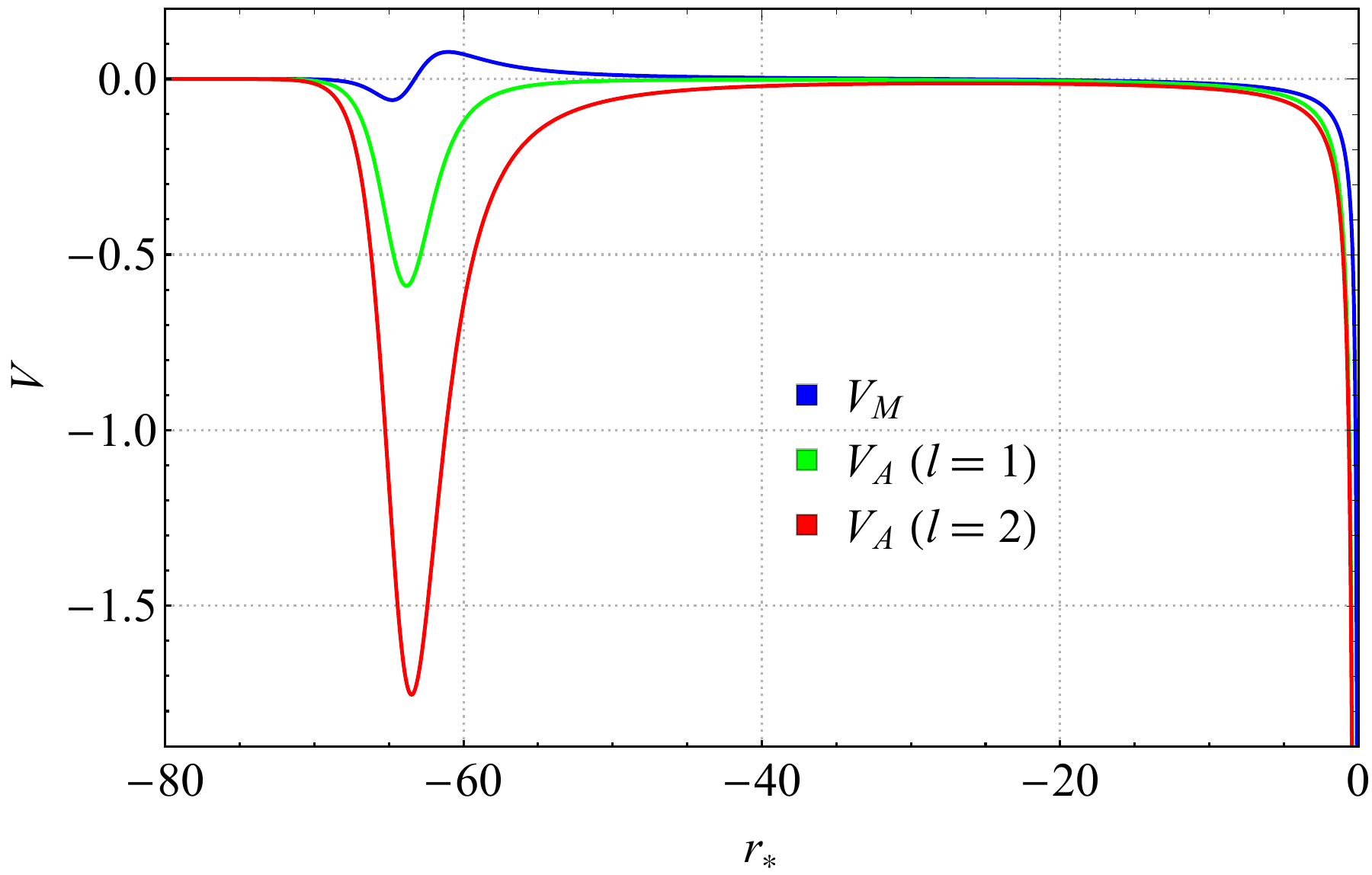}
	\caption{The effective potential $V(r_\ast)$ for monopole (blue) and axial-vector perturbations with $l=1$ (green) and $l=2$ (red). The parameter setting is $\mu = 1/10$ and $r_\pm=60$.}
	\label{fig:pot}
\end{figure}

\subsection{Monopole mode}

Evaluating \eqref{eq:near pole general omega} with the coefficients corresponding to the monopole potential, we obtain the following approximation for the trapped QNM spectrum:
\be \label{eq:near pole monopole omega}
\omega^2= -\frac{\mu^4r_-^2}{36 (n+1)^2}+\frac{\mu^2(4r_--3)}{6r_-}+\frac{(r_--1)(2r_--3)}{r_-^4} \,.
\ee
As explained, we expect this formula to be valid for modes that are sufficiently localized in the vicinity of the pole. Since the monopole potential exhibits a finite negative well when expressed in terms of the tortoise coordinate (see Fig.~\ref{fig:pot}), we anticipate that Eq.\ \eqref{eq:near pole monopole omega} will only correctly reproduce the spectrum of sufficiently unstable modes. The self-consistency of the approximation therefore requires that $\mu^2r_-\gg n+1$, so that
\be
\omega\simeq \frac{\mu^2r_-}{6(n+1)}i \,.
\ee
This predicts, in particular, a linear scaling of ${\rm Im}\,\omega$ with the pole distance $r_-$, in perfect agreement with the exact numerical results.

\subsection{Axial-vector mode}

Applying \eqref{eq:near pole general omega} to the axial mode potential we find
\be \label{eq:near pole axial omega}
\omega^2=V_0-\frac{1}{(2n+1)^2}\left[\frac{l(l+1)}{r_+}+\frac{\mu^2r_+}{3}-\frac{4r_+-5}{4r_+^2}\right]^2 \,,
\ee
with
\be\begin{aligned}
	V_0&=-\frac{(4r_+-5)l(l+1)}{2r_+^3}+\left(\frac{2}{3}-\frac{1}{2r_+}\right)\mu^2 \\
	&\quad +\frac{176r_+^2-496 r_++323}{48r_+^4} \,.
\end{aligned}\ee
The potential also exhibits a local minimum in this case, with the difference that $V_{\rm min}$ strongly depends on the multipole number $l$, explicitly $V_{\rm min}\propto -l(l+1)$; see Fig.~\ref{fig:pot}. A necessary condition for formula \eqref{eq:near pole axial omega} to be applicable is therefore that $\omega^2$ be sufficiently large and negative.

Two interesting regimes may be identified. The first is when $l=\mathcal{O}(1)$, $0<\mu\lesssim 1$ and $\mu^2r_+ \gg 2n+1$, so that
\be
\omega\simeq \frac{\mu^2r_+}{6(n+1/2)}i \,.
\ee
This is very similar to the result for the monopole mode. The second case is when $n=\mathcal{O}(1)$, $\mu\lesssim 1$ and $l(l+1)\gg r_+^2$, leading to
\be
\omega\simeq \frac{1}{2n+1}\left[\frac{l(l+1)}{r_+}+\frac{\mu^2r_+}{3}+\frac{n(n+1)(4r_+-5)}{r_+^2}\right]i \,.
\ee
Note that this regime is consistent, because although the potential well $V_{\rm min}$ also becomes deeper as $l$ increases, it does so slower than $\omega^2$. As mentioned in the main text, the last result explains the scaling with $l$ observed in the exact numerical calculations.

\bibliographystyle{apsrev4-1}
\bibliography{reference}

\end{document}